%% file: health_query_agent.tex
\documentclass[11pt]{article}

\usepackage[dvipsnames,table]{xcolor}
\usepackage{csquotes}
\usepackage[hyphens]{url}
\usepackage{hyperref}
\usepackage{standalone}

\usepackage{fontawesome5} % icons
\usepackage{tikz}

\usepackage{graphicx}
\usetikzlibrary{shapes.geometric, arrows.meta, positioning, fit, backgrounds, calc} 

\newcommand\shortUrl[1]{%
  \href{http://#1}{\nolinkurl{#1}}%
}
\usepackage{enumitem}
\setlist[itemize]{itemsep=0pt, topsep=5pt, parsep=0pt}
\setlist[enumerate]{itemsep=0pt, topsep=5pt, parsep=0pt}

\usepackage[most]{tcolorbox} % For the yellow background box!

\definecolor{cmblue1}{RGB}{239,243,255} % very light
\definecolor{cmblue2}{RGB}{189,215,231}
\definecolor{cmblue3}{RGB}{107,174,214}
\definecolor{cmblue4}{RGB}{49,130,189}
\definecolor{cmblue5}{RGB}{8,81,156}    % darkest (but still readable)

\definecolor{mainbg}{RGB}{255,250,235}   % Light Yellow
\definecolor{subbg}{RGB}{242,253,253}    % Very Light Teal
\definecolor{agentborder}{RGB}{180,180,180}
\definecolor{codeblue}{RGB}{0,50,130}
\definecolor{failred}{RGB}{140,40,40}
\definecolor{mygray}{RGB}{80,80,80}
\definecolor{inputfill}{RGB}{255,236,255} % Approx magenta!10-17ish
\definecolor{inputborder}{RGB}{200,100,200}

\usepackage{booktabs}
\usepackage{multirow}

\usepackage{authblk}

\usepackage[final]{acl}

\usepackage{times}
\usepackage{latexsym}

\usepackage[T1]{fontenc}
\usepackage[utf8]{inputenc}

\usepackage{microtype}

\usepackage{inconsolata}

\usepackage{graphicx}

\title{ClinQueryAgent: A Conversational Agent for Population Health Management}

\author{ \bf
Joseph S. Boyle$^{1,2,3}$,
Anthony Dranfield$^{3}$,\\
\bf Mike O'Neil$^{3}$,
\bf Maria Liakata$^{2,4}$,
Alison Q. Smithard$^{1,3,4}$
\\ \\
$^1$Canon Medical Research Europe $^2$Queen Mary University of London \\
$^3$Nottingham and Nottinghamshire ICB, NHS England \\
$^4$University of Edinburgh $^5$The Alan Turing Institute
}
\begin{document}

% load results values
\include{results_development}
\include{results_ziletti}
% \include{results_brand_names} % old
\include{results_brand_names2}
\include{results_real_eighty_two}

\include{results_user_statistics}

\maketitle
\begin{abstract}
% What does this paper add
In this paper we introduce \textsc{ClinQueryAgent}, a system for translating natural language population health questions into executable database queries using agents with access to both local and external knowledge bases.
Our novel architecture enables the use of powerful cloud-based language models whilst ensuring that no patient data leaves the secure environment. To combat inaccuracies over the course of longer dialogues due to context rot, information retrieval is delegated to a sub-agent.
We deploy the system via a chat window embedded within an existing population health management platform where it has been used by \totalUsers{} staff from \totalPractices{} healthcare practices covering a total of \totalPatientsInPractices{} patients in the UK's National Health Service (NHS).
We evaluate the system's capacity to autonomously handle a range of health informatics tasks on a constructed dataset and via a beta-testing phase.
Our results show that both analysts and clinicians are able to easily generate actionable information from patient health records using natural language requests requiring no programming expertise to verify. We make a public demo of the system available \footnote{Demo:  
\url{https://demo-899965260288.europe-west1.run.app/}}

\vspace{-1em}
\end{abstract}

\begin{figure}[htbp]
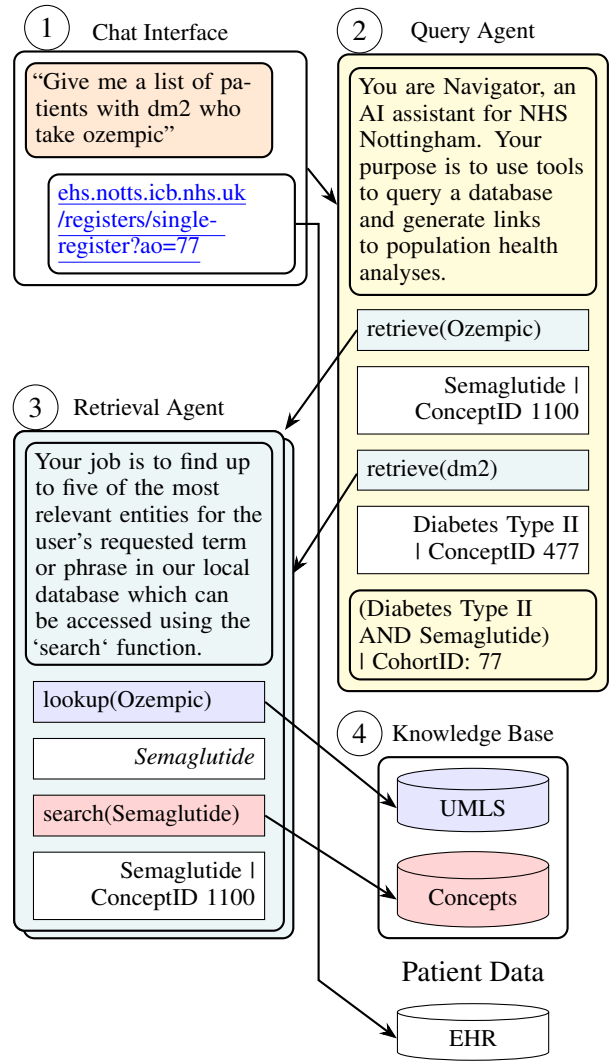

\centering
\includestandalone{figure_one}
\caption{
\colorbox{yellow!17}{ClinQueryAgent} parses \colorbox{orange!17}{natural language} \colorbox{orange!17}{questions} into database queries in an agentic loop. 
The Query Agent delegates the task of finding concepts to a \colorbox{teal!10}{Retrieval Agent} so that it can focus on creating the database query.
The Retrieval Agent has access to a local database of medical \colorbox{red!17}{concepts}, and the open-access ontology \colorbox{blue!10}{UMLS}, which is used to help the retrieval agent parse unusual acronyms and synonyms.
Diagram shows ClinQueryAgent's response to a user request about patients with type II diabetes mellitus (``dm2'').
}
\label{fig:question-to-query}

\end{figure}

\section{Introduction}

% focus has been elsewhere
Healthcare organisations collect large quantities of data, which clinicians can use to proactively manage the needs of their patients. However, the task of interrogating recorded healthcare data is complex; to perform sophisticated queries, clinicians typically rely on specialist data analysts.

% Ensure this is rendered on page 2.

% \begin{figure*}[t]
%   \includegraphics[width=\linewidth]{content/system_diagram3.png}
% 	\caption{System diagram showing a subset of QueryAgent's tools. The agent must enact multi-step plans using a concept database and a medical knowledge base to create an analysis query, which is then passed to the user who will be able to view the retrieved results. To ensure privacy, the agent's tools cannot access the Patient EHR database. The user may interactively refine the query by conversing with the agent, for instance prompting it to modify its results or asking follow up questions.}
% 	\label{fig:system-diagram}
% \end{figure*}

\begin{table*}[h]
\small
\centering
\resizebox{\textwidth}{!}{
	\begin{tabular}{p{5.5cm}p{8.5cm}}
    	\textbf{Workflow description} & \textbf{Real-world Example User Request} \\ 
		\specialrule{0.4pt}{0.25em}{0.5em}
        List each patient in a cohort & [General Practitioner] \textbf{``List the new hypertension diagnoses in} \\
        (returns a list) % ; \hyperref[fig:workflow-one]{example in Figure~\ref*{fig:workflow-one}}) 
            & \textbf{ the last 12 months''}\\
         & \textit{Intention: Clinical audit}\\
        \specialrule{0.4pt}{0.25em}{0.5em}
        Summarise age and sex statistics across a patient cohort & [Pharmacist] \textbf{``number of patients on semaglutide''}\\
        (returns the count, mean age, and sex) % \newline \hyperref[fig:workflow-two]{distribution; example in Figure~\ref*{fig:workflow-two}})
            & \textit{Intention: Check financial spend [Follow-up: Check if budget is targeting desired patient population e.g. high deprivation patients who cannot afford medication]}\\
        \specialrule{0.4pt}{0.25em}{0.5em}
        Show distribution of a variable for a cohort & [Mental Health Commissioning Manager] \textbf{``Can you find out the}\\
        (returns a histogram) %; \hyperref[fig:workflow-three]{example in Figure~\ref*{fig:workflow-three}})
            & \textbf{weights of the SMI QOF register patients excluding remissions?''}\\
        & \textit{Intention: Profile population of patients with mental illness to understand determinants of health e.g. obesity} \\
        \specialrule{0.4pt}{0.25em}{0.5em}
        Show the prevalence of a concept across practices / areas / districts & [Analyst] \textbf{``smoking statistics by district''} \\
        Optionally, compute the prevalence within a subpopulation, e.g. those with lung disease & \textit{Intention: Assess regional variation, plan smoking cessation services}\\
        (returns a percentage) % ; \hyperref[fig:workflow-four-view-one-large]{example in Figure~\ref*{fig:workflow-four-view-one-large}})
            & \\
        \specialrule{0.4pt}{0.25em}{0.5em}
        Show a To-Do list of patients for whom an & [General Practice Manager] \textbf{``aged 75 or older with 15+}\\
        intervention is required & \textbf{medications without a structured medication review''}\\
        (returns a list) % ; \hyperref[fig:workflow-five]{example in Figure~\ref*{fig:workflow-five}})
            & \textit{Intention: Schedule medication reviews} \\[0.5em]
        \specialrule{0.4pt}{0.2em}{-0.7em}
	\end{tabular}
}
\caption{
The five categories (workflows) of data analysis that may be handled and displayed by eHealthScope, with corresponding real-world user request examples.
The agent implicitly classifies the user request into a workflow description, which it typically describes at the beginning of its answer process: \textit{`To answer this question I will first search for hypertension and then create a cohort from those patients, and then link you to a register of those patients'.}
`SMI QOF register' refers to a specific concept for patients with one or more Serious Mental Illness.
}
\label{tab:workflows-example-requests}
\end{table*}

% we generate population health database queries from user questions, enabling clinicians to  using natural language data. ClinQueryAgent automates the databasing aspect of existing population health workflows, shown in Table~\ref{tab:workflows-example-requests}, time to perform these workflows from roughly 1-5m to 5-10s.

To answer population-level questions, it is necessary to identify the patient cohort(s) of interest, in a similar manner to identifying a target cohort in a research study \cite{duke_ohdsi_2018}. In Figure \ref{fig:question-to-query}, the required cohort comprises all patients with diabetes mellitus type II who are prescribed Ozempic. Patient cohorts can be identified by defining one or more ``concepts'' (sets of medical codes) and writing rules to describe eligible assignment patterns in the patient records (usually boolean logic, sometimes in combination with time constraints), then selecting all patients in which this assignment pattern is present. Many open source concept/cohort definition libraries exist such as \citet{thayer_creating_2024}. In this case, NHS Nottingham and Nottinghamshire ICB has defined 16,452 local concepts tailored to their local data format and analytical requirements; the above cohort could be identified by finding patients with the local concepts Diabetes II \emph{and} Semaglutide concepts (see Appendix \ref{app:example-concept-definitions}).

In this work, we introduce \textsc{ClinQueryAgent} -- an agent-based system for querying clinical databases through a task-oriented dialogue with a clinical user using generative language models \cite{bengio_neural_2003, sutskever_sequence_2014, radford_improving_2018}. Our system automates the creation of the population health queries, based on the observation that queries are easy to verify but labour intensive to produce. Given a natural language request, the system must express the user's goal as a logical query in terms of clinical concepts. Unlike SQL, a logical representation is easily understood by clinicians, empowering them to audit and co-author the query with the agent. This is subsequently compiled into an executable SQL query and run on patient data by the agent's host system.
%by NHS Nottinghamshire's eHealthScope application which hosts our agent.

\begin{figure*}[tbp]
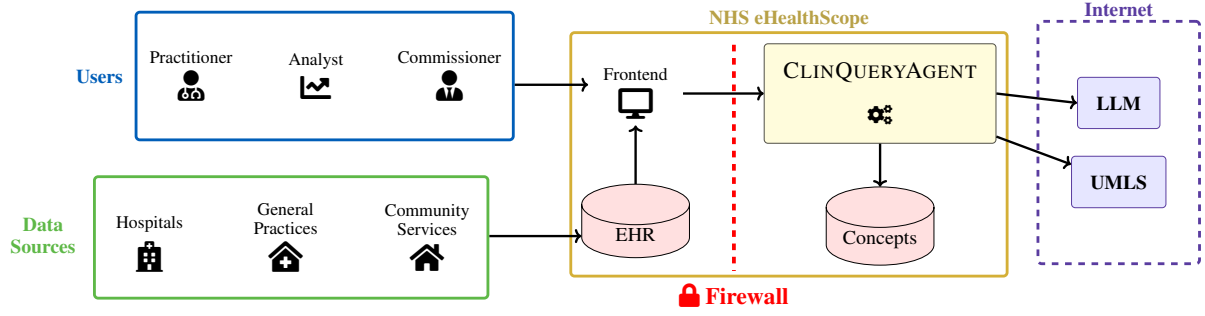

	\centering
	\includestandalone[width=\textwidth]{figure_two}
	\caption{
ClinQueryAgent (CQA) is situated within the eHealthScope application which serves a variety of users via the secure NHS intranet. Patient data is aggregated from healthcare centres into a central data warehouse (EHR) which is inaccessible to CQA by design, represented by the firewall between them.
}
	\label{fig:system-diagram}
    \vspace{-1.1em}
\end{figure*}
%\subsection{\textsc{ClinQueryAgent}}
\noindent{}We make the following contributions:
\begin{enumerate}
\vspace{-0.5em}
\item We demonstrate that conversational agents can provide fast and accurate natural language interfaces to clinical databases by evaluating ClinQueryAgent in a real world system used for population health management (here NHS Nottingham and Nottinghamshire ICB's population health analysis platform eHealthScope). 
% Automated query generation takes approximately 15 seconds to create queries derived from a database of more then 10,000 clinical concepts. Usage statistics indicate that the majority of queries are executed, indicating that they acceptably operationalise the user's intent.

\item Our system design is privacy-preserving since the agent is isolated from the patient data, showing how to securely deploy systems in environments where the underlying data is sensitive whilst taking advantage of frontier generative models deployed in the cloud.

\item We observe accuracy degrades and cost increases during multi-turn conversations due to `context rot' from large numbers of database results, and propose to reduce the main agent's context via delegation of database search to a subordinate agent.

\item We show that providing access to a knowledge base improves accuracy for queries involving specialist medical terminology.

\end{enumerate}

\section{Related Work}

Our proposed system sits at the intersection of two prevailing research topics: task-oriented dialogue and text-to-query systems.

\paragraph{Task-Oriented Dialogue}
% todo: make more relevant to our work.
In task-oriented dialogues (TODs), an agent and user collaborate to enact a user's goal \cite{yi_survey_2025}.

LLM agents are augmented with tools which enable them to take actions within an environment e.g. retrieving information from Wikipedia in order to obtain an answer. An early example was
SimpleTOD, a GPT-2 based agent which performs all aspects of task completion including database calls \cite{hosseini-asl_simple_2022}. Subsequent systems have continued to function via autoregressive language modelling, interleaving reasoning traces with task-specific actions. The so-called `ReAct' paradigm introduced by \citet{yao_reac_2023} introduced task delegation to external tools by issuing \emph{tool calls}.
% The ToolFormer paper \cite{schick_toolformer_2023} expanded this idea to a generic formulation for calling any tool of interest via API calls.
Using this paradigm, modern LLMs like Claude Code, Gemini-CLI, and OpenAI Codex are capable of autonomously performing complex programming tasks \cite{kwa_measuring_2025}.

% \begin{figure*}[h]
%     \small

%     \begin{tcblisting}{
%   listing only,
%   colback=yellow!17,
%   colframe=black!66!white,
%   sharp corners % Or use 'rounded corners'
% }
% #### **3. To get a report showing the frequency or distribution of a particular concept category (e.g. diagnoses, medications, ages) within a cohort:**

% Use the **Profiling tool (Report Mode)**.

% * **Workflow:**
%   1. `search(..., mode="concepts")`
%   2. `create_patient_cohort` (can be all active patients, if no specific cohort is requested)
%   3. `search(..., mode="concept_categories")`
%   4. `create_profile_url` (pass `denominator_cohort_id` and `concept_category_id`)
% \end{tcblisting}
% % \end{verbatim}
%     \caption{Prompt for workflow  3 in Table~\ref{tab:workflows-example-requests}, explaining how to compute distribution of a variable within a patient population. In eHealthScope, patient status is ``active'' if they are living. Since the tool \texttt{create\_profile\_url} can take different combinations of arguments, this workflow describes which arguments to pass.
%     }
% 	\label{fig:sample-workflow-template}
% \end{figure*}

\paragraph{Medical Text-to-Query}

Text-to-Query is the task of mapping a question to an executable query. Early medical text-to-query approaches used bespoke grammars to perform this task \cite{woodyard_natural_1981} and contemporary approaches use neural language models.
Recent applications include: querying synthetic records \cite{lee_ehrsql_2022,sivasubramaniam_sm3-text--query_2024, jiang_medagentbench_2025}; generating epidemiological analyses in real-world databases \cite{ziletti_generating_2025, moller-grell_agentic_2025}; and drafting implementations of inclusion criteria for medical studies \cite{yuan_criteria2query_2019, park_criteria2query_2024}. These approaches leverage techniques like in-context learning and vector-based retrieval augmented generation (RAG) \cite{brown_language_2020, lewis_retrieval-augmented_2021} to improve accuracy.

% Other agents such as Claude Code \cite{anthropic_anthropicsclaude-code_2026} used straightforward lexical search for information retrieval.

\section{Method}
% A method is like a recipe.
ClinQueryAgent is an LLM agent system. This section describes the prompt design process, tool use capabilities, and sub-agent delegation mechanisms.

\subsection{Prompt design}

To guide our agent on how to answer different types of questions, we provide templates in the ClinQueryAgent's system prompt to demonstrate which tools to call in order to handle a user's request. This paradigm is inspired by in-context learning which is a standard method to improve the task-specific performance of LLMs, by providing training examples in the prompt. However, unlike in-context learning examples, templates do not reference specific entities, which avoids biasing the generation process towards including these entities (in our case concepts) even when they are irrelevant~\cite{schulhoff_prompt_2024}.

In this case, our agent is serving the eHealthScope application, thus we create templates for the 5 categories (``workflows'') of data analysis that may be handled by eHealthScope (see Table~\ref{tab:workflows-example-requests}).
%These workflows are applicable to population health management systems in general, for example querying patients and summarising cohort statistics.

%The full prompt is shown in Appendix~\ref{app:prompt}.

\subsection{Tools}
ClinQueryAgent is provided with tools to perform:
\begin{enumerate}
    % \item \textbf{concept retrieval} via a sub-agent
    \item \textbf{lexical concept search} within the NHS database; this tool fetches concepts which contain (sub)strings specified by the agent
    \item \textbf{patient cohort creation} from query logic; this tool does not return any value to the agent, but converts the query logic provided by the agent to executable SQL, then runs the SQL and caches the resulting patient cohort in the eHealthScope database
    % to create a SQL query from a logical AND/OR query
    \item \textbf{url creation} to results display; this tool returns a url which the user may click to see the specified patient cohort displayed in the eHealthScope UI
\end{enumerate}

% \noindent{}Tools were implemented in the industry standard JSON tools format.
% We found that the agent's performance was heavily contingent on the quality of the provided tool documentation.
% %We describe the tools further in Appendix~\ref{app:tool-definitions}.

\subsection{Accessing external knowledge}

To help the agent retrieve concepts not well defined within its pre-trained knowledge, there is the option to provide additional tools to retrieve information from knowledge bases. 
% We experiment with providing access to the Unified Medical Language System® (UMLS®), a  compendium of many ontologies, containing thesauri of synonyms for each concept as well as the labelled relations between concepts.
We experiment with providing access to the OHDSI standardised vocabularies, containing over 10 million codes \cite{reich_ohdsi_2024}, via a \textbf{lookup} tool.
The vocabularies contain a rich mapping of chemical compounds to drug names, which enables the agent to link e.g. `ozempic' to `semaglutide', and extend data from the Unified Medical Language System (UMLS).
%This improves the agent's success on tasks involving niche knowledge, as shown in Figure~\ref{fig:question-to-query}.

\subsection{Sub-Agent delegation of information retrieval}
Identifying appropriate medical concepts in a real-world clinical database is a challenging task, involving elements of semantic parsing and disambiguation, handling of polysemy, synonymy, metonymy, and abbreviations, as well as local knowledge on naming conventions. 
The lexical concept search tool 
% As described above, we provide a search tool for performing a lexical substring database search. This tool 
may return a large number of concepts (`diabetes' returns 407 results, for instance, since there are many variant concepts relating to diabetes) which the agent must parse to determine which concepts are relevant to the user's query. Over the course of a multi-turn conversation the number of concepts in the context grows and a degradation in query accuracy is observed. We attribute this degradation to the long bloated context, as noted in prior work~\cite{liu_lost_2024, laban_llms_2025} and colloquially referred to as `context rot'.

\begin{figure}[!thbp]
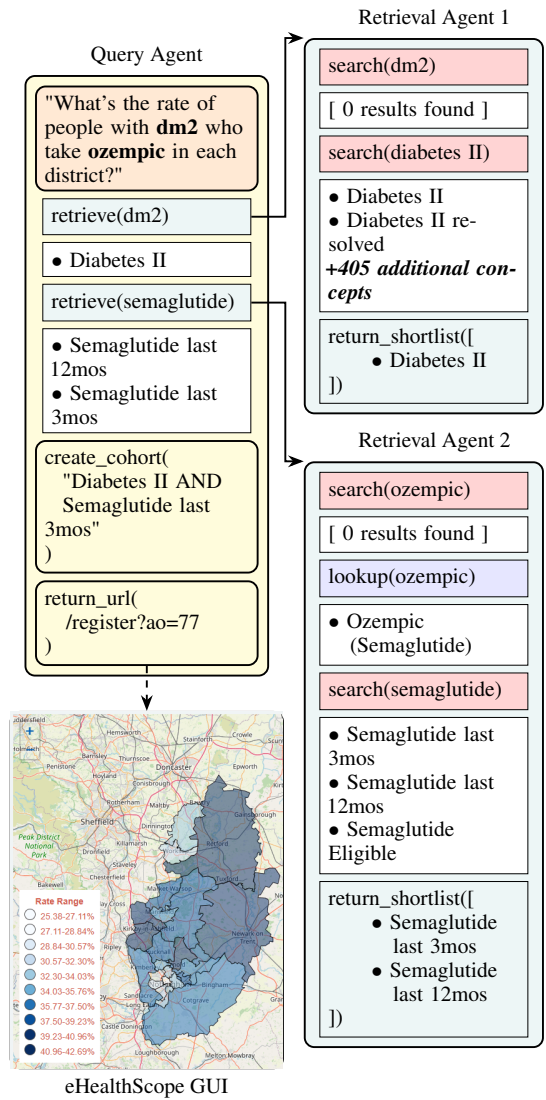

    \centering
    \includestandalone[scale=0.9]{figure_three}
    \caption{An example of computing statistics across regions and visualising them in the eHealthScope GUI. Linking to the most appropriate concept can be a complex multi-step task involving lookups, and filtering large quantities of results: for example the diabetes II search returns 400+ concepts.
    The entity linking task is delegated to sub-agents, which are created with a fresh context for each \colorbox{teal!17}{{retrieve(...)}} call. The QueryAgent determines which of the retrieved concepts to include in the query; here, patients with `Semaglutide last 3 mos'' have the most recent evidence of taking semaglutide.}
    \label{fig:figure_three}
\end{figure}

To reduce irrelevant context, we experiment with using a sub-agent to perform the search process.
In this configuration (see~Fig.~\ref{fig:figure_three}), the primary agent (``Query Agent'')
instantiates a new sub-agent (``Retrieval Agent'') via a delegation tool, prompting it with the search term. The retrieval agent performs its own iteration loop, exploring the concept database, eventually returning a small number of concepts and explaining their relevance, e.g. \textit{``Returned the concept for `Semaglutide' as it is the generic form of the requested term `Ozempic' ''}.

The sub-agent has two tools: \textbf{search} for searching for concepts, and \textbf{return\_shortlist}, for returning concepts alongside any explanation that the sub-agent wants to give as to why those concepts are relevant. Calling \texttt{return} ends the sub-agent's routine, and Query Agent continues with the newly retrieved concepts.

Use of a sub-agent keeps the Query Agent's context window shorter, allowing it to reason over a smaller and more relevant set of concepts. Additionally, each concept search triggers the instantiation of a new sub-agent, allowing each sub-agent's context to remain focussed on a single task.

%Appendix~\ref{app:prompt} shows the system prompt instructions for the sub-agent.

\section{Experimental Setup}

\paragraph{Model} We run a real world evaluation in a dialogue based deployment of ClinQueryAgent. Our framework is agnostic to the choice of LLM. We choose to evaluate using Gemini 2.5 Flash \cite{comanici_gemini_2025}, based on its frontier price-performance characteristics in agentic benchmarks  \cite{kapoor_holistic_2025}. 
The `thinking' token budget is configured to 0, as we found it to slow down response times substantially without improving quality. Greedy decoding is used (temperature=0). 

\paragraph{Datasets} First, we adapted a benchmark of epidemiological cohort questions from \citet{ziletti_generating_2025}'s \textit{EpiCoho} dataset to the context of our application.
Second, to test the agent's ability to leverage an external knowledge base, we created a synthetic dataset of 100 questions in the form ``List patients who take X'' where X is a brand name sampled at random from the UK medicines agency\footnote{\tiny \url{www.gov.uk/government/publications/category-lists-following-implementation-of-the-windsor-framework}}. These questions require matching of the brand name to its generic name concept e.g. Ozempic to semaglutide. Third, we manually reviewed each user request from our real-world beta-deployment and constructed a dataset comprising 82 questions. Appendix~\ref{app:data-curation} gives further details about dataset creation.

\paragraph{Tasks} We evaluated the agent's ability to answer a) Single Questions, where one agent is asked a single question, and b) Chained Questions, where all questions are asked sequentially within one chat.

\begin{figure}[!tbh]
    \centering
    \caption{
Cumulative message count for ClinQueryAgent in its beta deployment phase from 2025-06-26 – 2026-02-13. Clinical staff include GPs, care-coordinators, and pharmacists; Management and Commissioning includes operations staff, practice managers; Data \& Technology staff include population health analysts, software developers, and data management staff.
    }
    \includegraphics[width=\linewidth]
    {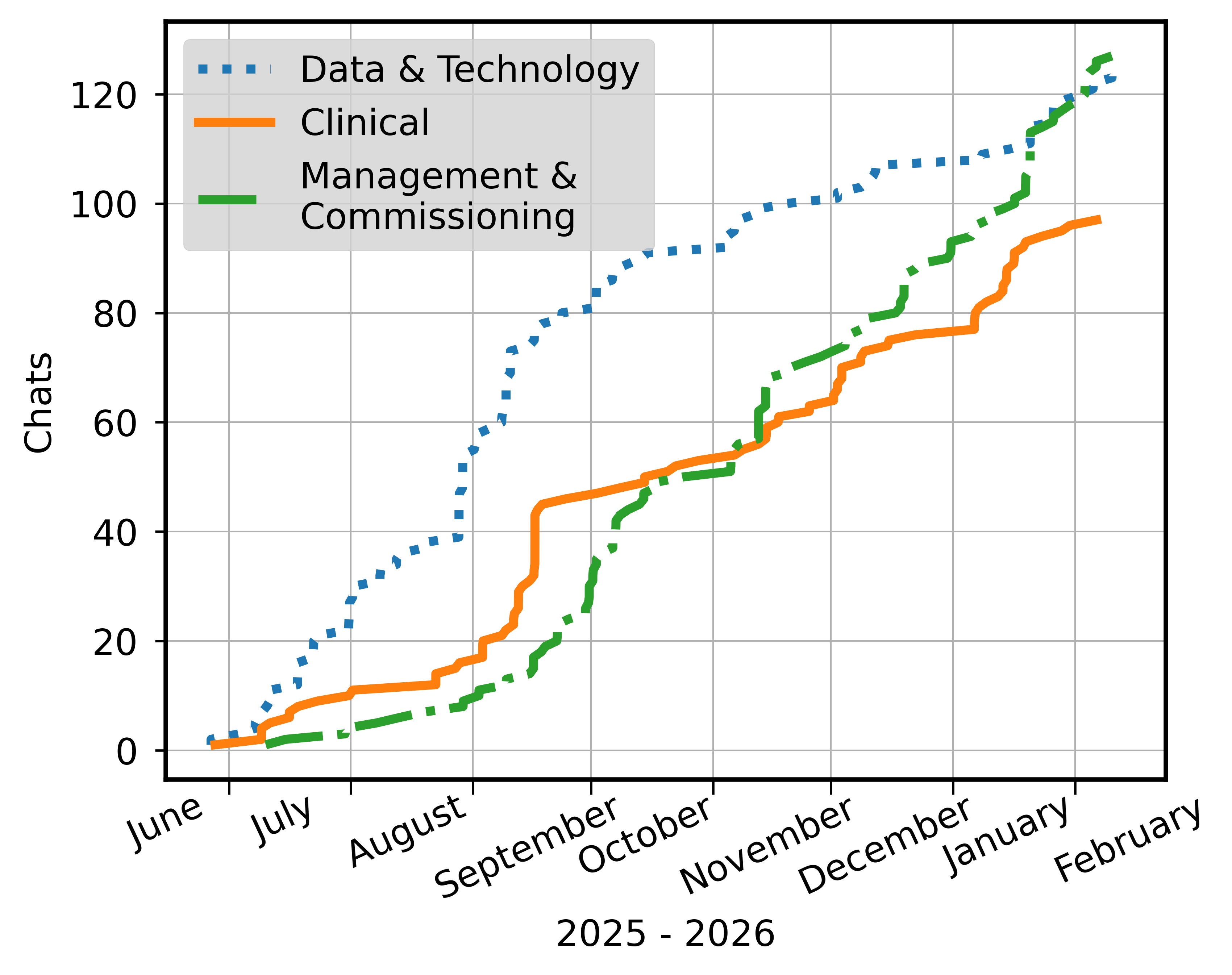}
\label{fig:cumulative-message-count}
\vspace{-10mm}
\end{figure} 

\paragraph{Metrics} The same question may be correctly answered with different queries (sometimes using different concepts), thus we do not evaluate the generated query verbatim against the reference query but assess similarity as follows: \emph{recall} is the proportion of the patients in the reference cohort who are in the retrieved cohort, \emph{precision} is the proportion of patients in the retrieved cohort who are in the reference cohort, and \emph{F1-score} is their harmonic mean.
We further characterise performance using the number of \emph{tokens} consumed, trajectory \emph{time}, and \emph{tool calls}. Each of these metrics is computed per trajectory and averaged.

\definecolor{lightgray}{gray}{0.95}

\begin{table*}[thb!]
\tiny
\centering
\resizebox{\textwidth}{!}{

    \begin{tabular}{@{}llcccrrr@{}}
    % \toprule
    \addlinespace[0.1em]
    & & \multicolumn{3}{c}{Patient Overlap} & \multicolumn{3}{c}{Mean Per Question} \\
    \addlinespace[-0.3em]
    \cmidrule(lr){3-5} \cmidrule(l){6-8}
    \addlinespace[-0.3em]
    Type & Method & R & P & F1 & Tokens & Time (s) & Tools \\
    \addlinespace[-0.3em]
    \midrule
    
    \multicolumn{8}{@{}l}{\textbf{EpiCoho-M Dataset}} \\
    %\addlinespace[0.2em]
\rowcolor{lightgray}  & QueryAgent          & \Recall               & \textbf{\Precision}           & \Fscore                   & \textbf{\Tokensmean}       & \textbf{\Elapsedtimeavg}       & \Toolinvocationsmean \\
\rowcolor{lightgray} Single  & + Retrieval Agent     & \irRecall             & \irPrecision                  & \irFscore                 & \irTokensmean     & \irElapsedtimeavg     & \textbf{\irToolinvocationsmean} \\
\rowcolor{lightgray}  & + Retrieval Agent + UMLS    & \textbf{\irumlsRecall} & \textbf{\irumlsPrecision}    & \textbf{\irumlsFscore}    & \irumlsTokensmean & \irumlsElapsedtimeavg & \irumlsToolinvocationsmean \\
    \addlinespace[0.2em]
       & QueryAgent          & \sRecall                      & \sPrecision           & \sFscore                  & \sTokensmean                  & \sElapsedtimeavg            & \textbf{\sToolInvocationsmeanpersample} \\
Chained& + Retrieval Agent     & \sirRecall                    & \textbf{\sirPrecision} & \sirFscore               & \sirTokensmean                & \textbf{\sirElapsedtimeavg} & \sirToolInvocationsmeanpersample \\
       & + Retrieval Agent + UMLS    & \textbf{\sirumlsRecall}       & \sirumlsPrecision      & \textbf{\sirumlsFscore}  & \textbf{\sirumlsTokensmean}   & \sirumlsElapsedtimeavg      & \sirumlsToolInvocationsmeanpersample \\
    
    \addlinespace[0.5em] % Large gap between datasets
    %\hline

    \multicolumn{8}{@{}l}{\textbf{UK Medicine Brands Dataset}} \\
    %\addlinespace[0.2em]
 
\rowcolor{lightgray}        & QueryAgent          & \bbRecall                     & \bbPrecision                  & \bbFscore                     & \textbf{\bbTokensmean}        & \textbf{\bbElapsedtimeavg} & \textbf{\bbToolinvocationsmean} \\
\rowcolor{lightgray}Single  & + Retrieval Agent     & \bbirRecall                   & \bbirPrecision                & \bbirFscore                   & \bbirTokensmean               & \bbirElapsedtimeavg & \bbirToolinvocationsmean \\
\rowcolor{lightgray}        & + Retrieval Agent + UMLS    & \textbf{\bbirathenaRecall}    & \textbf{\bbirathenaPrecision} & \textbf{\bbirathenaFscore}    & \bbirathenaTokensmean & \bbirathenaElapsedtimeavg & \bbirathenaToolinvocationsmean \\
    %\addlinespace[-0.1em]

    \addlinespace[0.5em]

    \multicolumn{8}{@{}l}{\textbf{Real World Dataset}} \\
    %\addlinespace[0.2em]
\rowcolor{lightgray}        & QueryAgent          & \rRecall                   & \textbf{\rPrecision}       & \rFscore                   & \rTokensmean                & \textbf{\rElapsedtimeavg}& \rToolinvocationsmean \\
\rowcolor{lightgray} Single & + Retrieval Agent     & \textbf{\rirRecall}        & \textbf{\rirPrecision}     & \textbf{\rirFscore}        & \textbf{\rirTokensmean}     & \rirElapsedtimeavg       & \rirToolinvocationsmean \\
\rowcolor{lightgray}        & + Retrieval Agent + UMLS    & \textbf{\rirumlsRecall}    & \textbf{\rirumlsPrecision} & \textbf{\rirumlsFscore}    & \rirumlsTokensmean          & \rirumlsElapsedtimeavg   & \textbf{\rirumlsToolinvocationsmean} \\
    \addlinespace[0.2em]
    \multirow{3}{*}{Chained}
      & QueryAgent          & \srRecall                 & \srPrecision                  & \srFscore                 & \srTokensmean             & \textbf{\srElapsedtimeavg}    & \textbf{\srToolinvocationsmean} \\
      & + Retrieval Agent     & \textbf{\srirRecall}      & \srirPrecision                & \srirFscore               & \textbf{\srirTokensmean}  & \srirElapsedtimeavg           & \srirToolinvocationsmean \\
      & + Retrieval Agent + UMLS    & \textbf{\srirumlsRecall}  & \textbf{\srirumlsPrecision}   & \textbf{\srirumlsFscore}  & \srirumlsTokensmean       & \srirumlsElapsedtimeavg       & \srirumlsToolinvocationsmean \\

    \hline
    \addlinespace[-0.4em]
    \end{tabular}
}
\caption{
    Performance metrics on the \textit{EpiCoho-M} epidemiological cohort questions, the Real World beta-test questions and the UK brands datasets, with/without the concept retrieval sub-agent, and with/without access to UMLS. We assess accuracy according to how closely the retrieved patient cohort matches the expected patient cohort in a real-world database, using recall (R), precision (P) and F1-score (F1). We also show the mean number of tokens used, time, and number of tool calls per question.
}
    \vspace{-1em}
\label{tab:epi-coho-results}
\end{table*}

\section{Results}

Figure~\ref{fig:cumulative-message-count} shows use of the real-world system during the beta-testing phase, with steady uptake during this period. The tool is most popular with technology staff which we ascribe to its use by population health analysts, whose core work is querying population-level digital data. Overall, there were \totalUsers{} unique users who engaged in \totalChats{} chats. User feedback is analysed in Appendix~\ref{app:additional-results}.

Table~\ref{tab:epi-coho-results} shows performance on the three datasets. We compare the performance of a naive single agent versus extension with the retrieval agent, with or without an external UMLS resource. All configurations achieve similar F1-scores in the single question setup for the EpiCoho-M and Real World datasets. Conversely in the chained setup a large degradation in accuracy is observed when QueryAgent is used without a retrieval agent, with the F1-score dropping from \Fscore{} to \sFscore{} on EpiCoho-M and from \rFscore{} to \srFscore{} on the Real World dataset. Using the Retrieval Agent reduces the number of tokens consumed from \sTokensmean{} to \sirTokensmean{} (↓93\%).
Surprisingly, F1-score on the Real World dataset improves from \rirumlsFscore{} on single questions to \srirumlsFscore{} on chained questions, suggesting that the context of previously answered questions improves accuracy.

Despite the accuracy benefits of sub-agent use, in many configurations the answer took longer to generate.
In the UK medicine brand names dataset, the UMLS agent took twice as long to construct queries (\bbirathenaElapsedtimeavg{}s vs \bbElapsedtimeavg{}s) but achieved a superior average F1-score: \bbirathenaFscore{} vs \bbFscore. This could be sped up via the parallelisation of sub-agents.

\section{Discussion}

In their overview of Agentic AI in healthcare, \citet{karunanayake_next-generation_2025} identify privacy  and explainability as key challenges for successful integration into real world clinical practice. We address these via \emph{structural privacy} and \emph{verifiable logic}, as follows.
By strictly separating patient data (execution layer) from medical concepts and queries (reasoning layer), we achieve a structural guarantee of privacy that exceeds standard GDPR/HIPAA compliant cloud integration e.g. Microsoft Azure; this design enables frontier model use and protects patient data.
% Since no patient data is accessed by the agent, this  we are able to use the latest and most powerful frontier models.
By generating the queries in logical form, e.g. "Diabetes 2 AND Semaglutide", we enable users to check the generated query. This shifts the question from `Do I trust this LLM?' to `Is this logic a correct implementation of my original request?' The dialogue interface to ClinQueryAgent allows a collaborative style of query creation in which the user helps to refine the draft through follow-up instructions.

Whilst the ClinQueryAgent framework was tailored to the eHealthScope population health management platform, the features of our system (multi-agent decomposition, logical query language, robust isolation of patient data) are generic and can be easily generalised, as has been shown in the public demo, which interrogates a public database and has a bespoke UI.

\section{Conclusion}
We demonstrate ClinQueryAgent, a novel task-oriented dialogue system for performing population health analysis.
Our evaluation shows that deploying a sub-agent for the task of information retrieval helps the system to maintain coherence over the course of long conversations and that integrating external medical vocabulary information augments performance on rare  medicine names.
ClinQueryAgent is integrated into NHS Nottingham and Nottinghamshire ICB's population health platform and has been trialled by \totalUsers{} clinicians from \totalPractices{} healthcare practices covering a total of \totalPatientsInPractices{} patients. Automated query generation takes approximately 15 seconds for queries drawing from a database of 16,452 clinical concepts, with usage data indicating that most queries acceptably operationalise the user's intent. We propose this design as an exemplar for future AI-enabled text-to-query systems where data security is a priority.

\section{Limitations}

The system was implemented in a proprietary database used for real world clinical care, and so it was necessary to modify the EpiCoho-M dataset. This means that we cannot directly benchmark our results against other methods which evaluate on this dataset.

ClinQueryAgent has some limitations in terms of unknown concepts or knowledge. Firstly, ClinQueryAgent can only use pre-existing local concepts, meaning that some questions cannot be answered, e.g. those relating to a particularly novel disease for which there is no concept yet. Future work could tackle the generation of new concepts at inference time, enabling more varied and precise analyses. Secondly, ClinQueryAgent does not have knowledge of local terminology; for instance our agent did not understand when a user asked for a type of document called a ``Diabetes quick''. In future, we could expand the knowledge base to include local documents.

There are also limitations on how reliably the system can answer sophisticated questions like ``Find me patients who should be on the End Of Life Register''. This question requires clinical judgement, which may or may not have already been codified into a concept(s) that is accessible to ClinQueryAgent.

\section{Ethics Statement}
This study incorporating a service evaluation uses anonymised, routinely collected data from National Health Service (NHS) Nottingham and Nottinghamshire services and does not require NHS Research Ethics Committee (REC) approval according to NHS Health Research Authority (HRA) guidelines\footnote{\tiny \url{www.hra-decisiontools.org.uk/research/}}. The need for consent to participate was deemed unnecessary according to the aforementioned HRA guidelines.

We have carefully designed the system to prevent the possibility of patient data being accessed by the agent.
To guarantee this, we create a selection of database tools which do not provide access to the patient data (`Patient' database table), or any tables derived from it. The agent is limited to generating a \textit{query}, which is executed by the user according to their respective level of permissions when they access the results URL, shown in Fig.~\ref{fig:question-to-query}.

% To minimise the risk of misinformation due to an error from the agent, we used a boolean query language comprising the operations AND, OR, AND NOT, which offers a simple and mechanistic explanation of the generated result.

J.B. and A.S. had honorary contracts with the NHS with de-identified access to the data in order to conduct this work, as part of the regular delivery of patient care under the remit of NHS Nottingham and Nottinghamshire Strategic Analysis and Intelligence Unit.
% which provide analyst-level access to the data, sufficient to develop and evaluate performance. There were no personal identifiers in the data.

%Ethically there is a responsibility to treat patient data as confidential and to share it strictly on a `need to know' basis.
%Concepts pertaining to a certain record would be considered confidential information, even if the patient name and other identifiers were pseudonymised
%\footnote{Under UK law, the information pertaining to a person is considered confidential even if it is anonymised: \\
% \tiny
%\shortUrl{digitalregulations.innovation.nhs.uk/regulations-and-guidance-for-developers/all-developers-guidance/proof-of-concept-using-anonymous-or-artificial-health-data/}

\section*{Acknowledgments}
We thank the NHS Nottingham and Nottinghamshire ICB System Analytics Intelligence Unit (SAIU) for their expertise and contributions to this project, particularly Anthony Dranfield and Ben Clarke for their contributions to the project, and Carl Davis for organising the collaboration. We thank Patrick Schrempf for reviewing this manuscript and the project's source code.  This work was supported by the Engineering and Physical Sciences Research Council [grant number EP/Y009800/1], through funding from Responsible AI UK (KP0016) as a Keystone project lead by Maria Liakata.
\vspace{-5mm}
\bibliography{references}

\clearpage
\appendix

\onecolumn
\clearpage

%%% Toggle these to control what is in the appendix. %%% 
\newif\ifShowGUIScreenshots
\newif\ifShowPrompts
\newif\ifshowToolDefinitions
\newif\ifShowAdditionalResults
\newif\ifShowDataCuration
\newif\ifShowCohortDefinitions

\ShowCohortDefinitionstrue
% \ShowCohortDefinitionsfalse

%\ShowGUIScreenshotstrue
\ShowGUIScreenshotsfalse

%\ShowPromptstrue
\ShowPromptsfalse 

% \showToolDefinitionstrue
\showToolDefinitionsfalse

\ShowAdditionalResultstrue
% \ShowAdditionalResultsfalse

\ShowDataCurationtrue
% \ShowDataCurationfalse

%%%%%%%%%%%%%%%%%%%%%%%%%%%%%%%%%%%%%%%%%

\ifShowCohortDefinitions
    \section{Example Concept Definitions}
    \label{app:example-concept-definitions}
    
% In NHS Nottingham, cohort definitions are termed `EventConcepts' and so we refer to them as `concepts'-- not to be confused with a SNOMED-CT concept which is just a code (e.g. 44054006 | Type II Diabetes Mellitus); elsewhere in the literature you may see concepts referred to as `phenotypes' \cite{thayer_creating_2024}. Cohort definitions comprise 1) one or more medical codes 2) some logic about how to select patients based on those medical codes - an `inclusion' criteria - and 3) an optional set of exclusion criteria.
    
    \begin{table}[h]
        \centering
        \small
        \vspace{1.25mm}
    
    \begin{tabular}{|p{2.25cm}|p{1.8cm}|p{3.75cm}|p{5.5cm}|}
    \hline
         \textbf{Concept} & \textbf{Code ID} & \textbf{Code Description} & \textbf{Pseudocode} \\
    \hline
           & 44054006  & Type II Diabetes & \small \texttt{SELECT pat\_id} \\
         \textbf{Diabetes}   & 472969004 & History of diabetes type 2 & \small \texttt{FROM Conditions} \\
        \textbf{type II}     & 443694000 & Uncontrolled type 2 diabetes & \small \texttt{WHERE code IN (} \\
                   &     ...   & \hspace{1cm} \textit{+160 more} & \small \texttt{\hspace{0.5cm} 44054006, 472969004, ...)} \\
    \hline
            & 777514008  & Semaglutide only product & \small \texttt{SELECT pat\_id} \\
        \textbf{Semaglutide} & 782102009  & Semaglutide 1.34 mg/mL & \small \texttt{FROM Medications} \\ 
        \textbf{- last 3 months} &            & solution for injection & \small \texttt{WHERE code IN (777514008, ...)} \\
                       &     ...   & \hspace{1cm} \textit{+10 more} & \small \texttt{AND date $\ge$ DATE\_SUB(NOW, 3M)} \\
    \hline
    \end{tabular}
        \caption{Example NHS Nottingham and Nottinghamshire ICB concept definitions. The concept codesets are described in columns 2 \& 3, and the final ``Pseudocode'' column illustrates the rules to combine them; these are usually boolean logic, sometimes including time constraints. The corresponding patient cohort comprises all patients with this code assignment pattern present in their record.
    }
        \label{tab:concept-as-multi-vocab-codelist}
    \end{table}
\fi

\ifShowGUIScreenshots
    \section{eHealthScope UI Data Displays}
    \label{app:ehealth-scope-displays}
    Below we show 5 GUI screenshots corresponding to the 5 example requests in Table~\ref{tab:workflows-example-requests}.
    \vspace{-0.5mm}
    
    \begin{figure*}[!htbp]
        \centering
        \includegraphics[width=\linewidth]{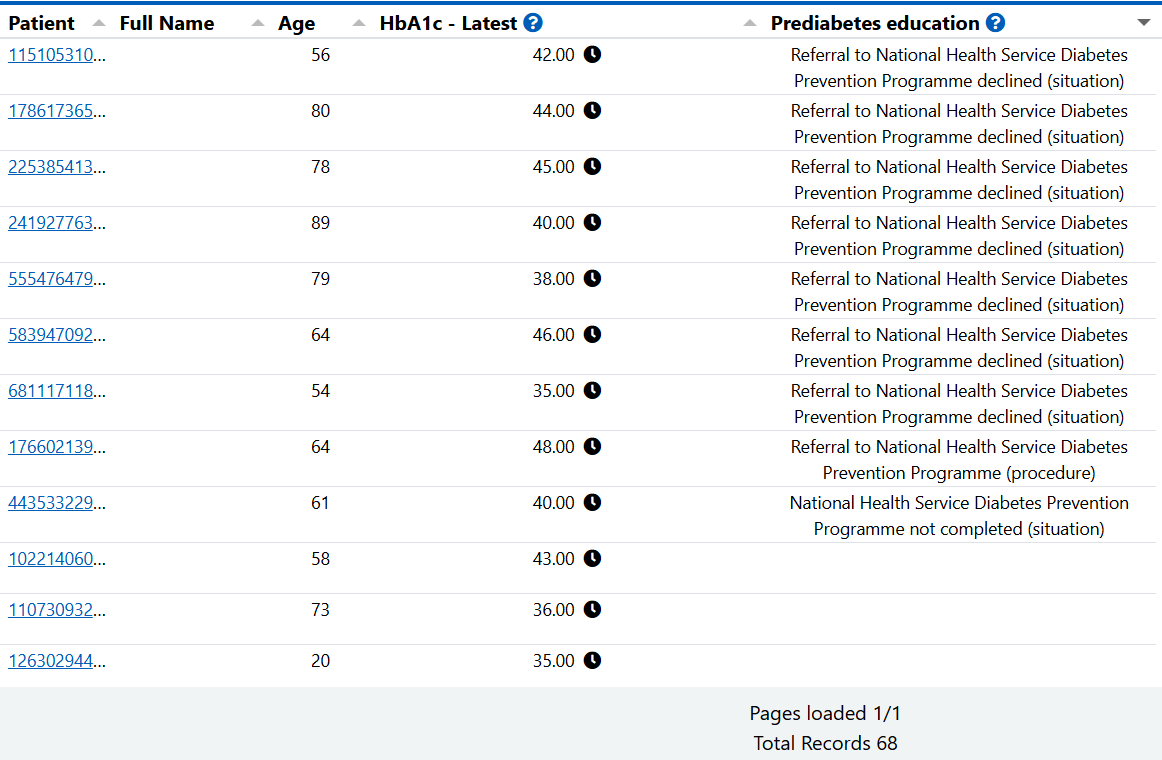}
        \caption{\textbf{Workflow 1:} ``List the new hypertension cases in the last 12 months''}
        \label{fig:workflow-one}
    \end{figure*}
    \vspace{-0.5em}
    
    \begin{figure*}[!htbp]
        \centering
        \includegraphics[width=1\linewidth]{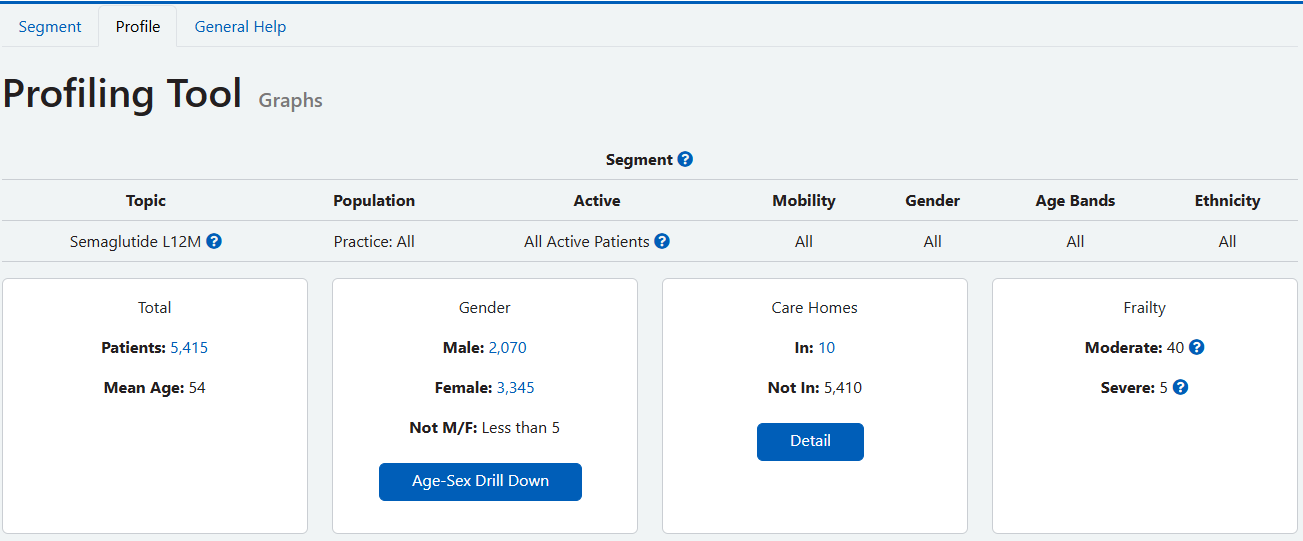}
        \caption{\textbf{Workflow 2:} ``number of patients on semaglutide''}
        \label{fig:workflow-two}
    \end{figure*}
    
    \begin{figure*}[!htbp]
        \centering
        \includegraphics[width=1\linewidth]{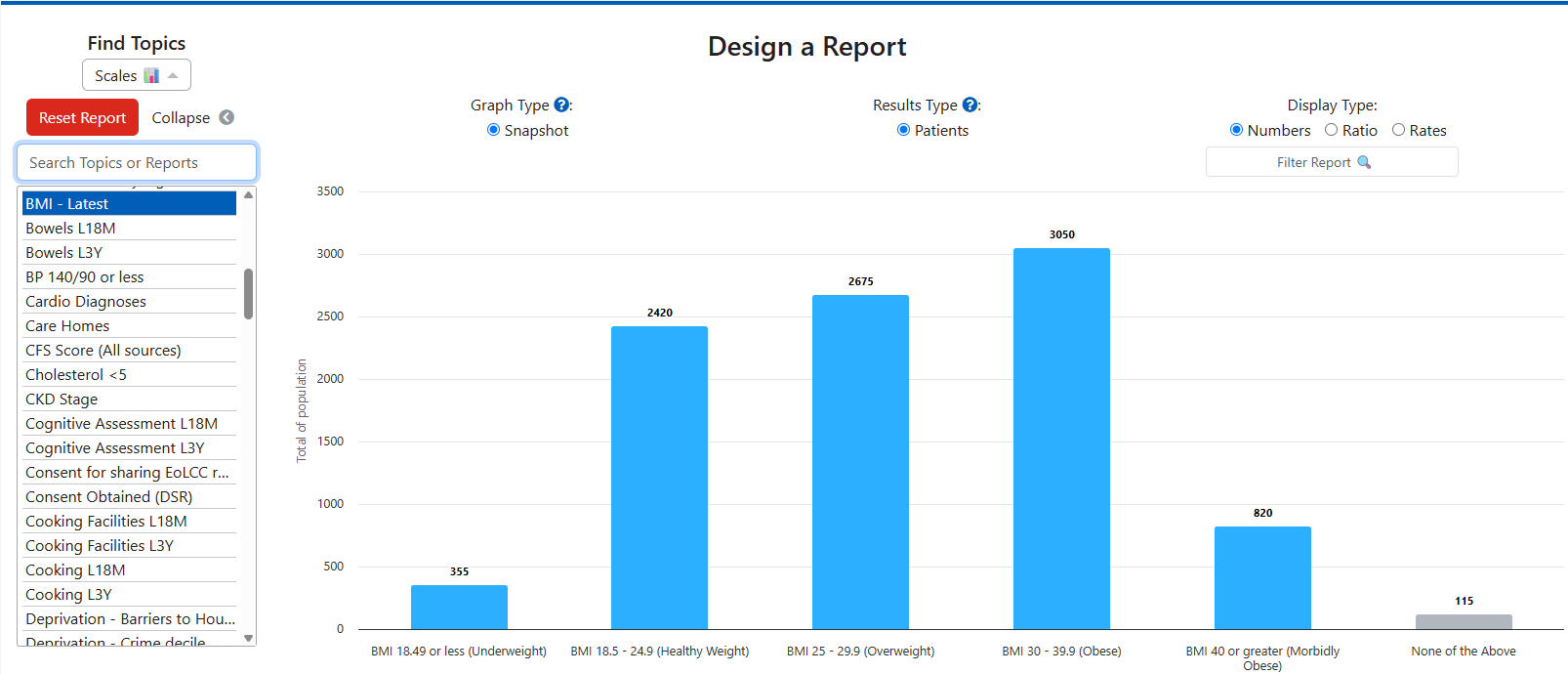}
        \caption{\textbf{Workflow 3:} ``Can you find the weights of patients on the [Serious Mental Illness Quality Outcomes Framework register] excluding remissions?''}
        \label{fig:workflow-three}
    \end{figure*}
    
    \begin{figure*}[!htbp]
        \centering
        \includegraphics[width=1\linewidth]{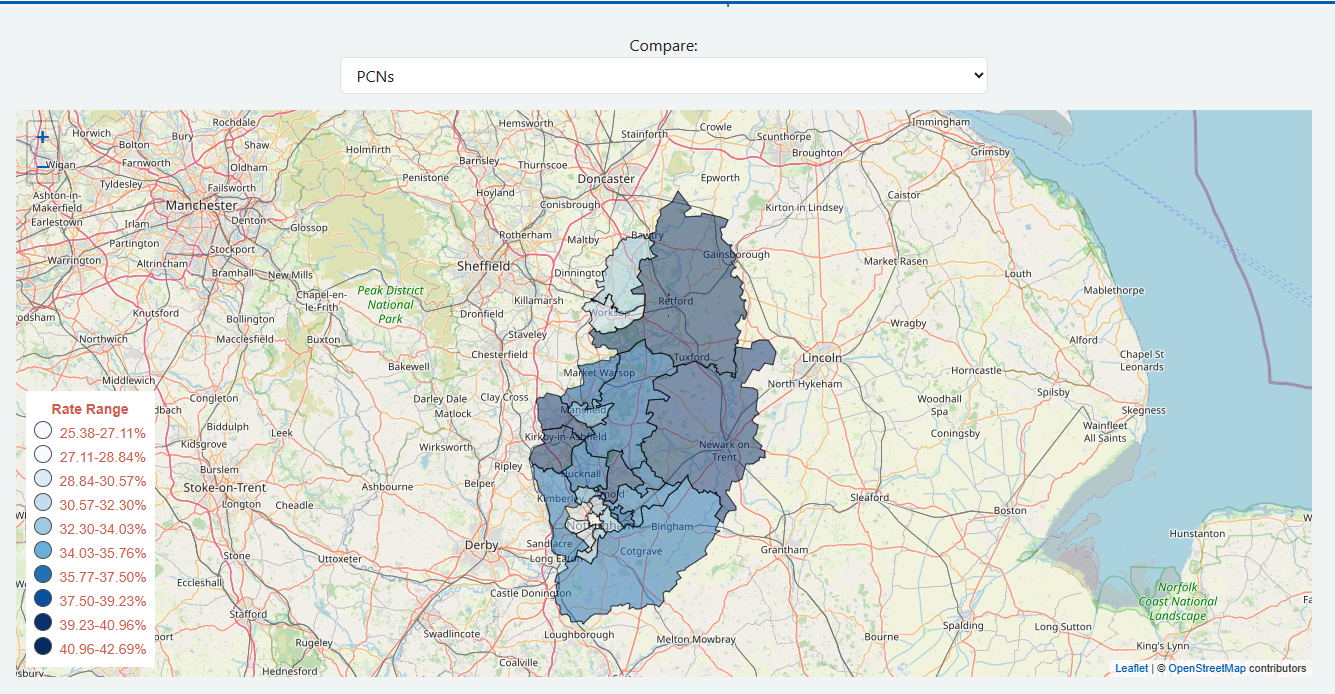}
        \caption{\textbf{Workflow 4:} ``Smoking statistics by district''. Results may be viewed as a bar chart of percentages, or as an interactive map within eHealthScope.}
        \label{fig:workflow-four-view-one-large}
    \end{figure*}
    % \begin{figure}
    %     \centering
    %     \includegraphics[width=0.25\linewidth]{content/workflow_four_view_one.png}
    %     \caption{``Smoking statistics by district''}
    %     \label{fig:workflow-four-view-one}
    % \end{figure}
    
    \begin{figure}[!htbp]
        \centering
        \includegraphics[width=1\linewidth]{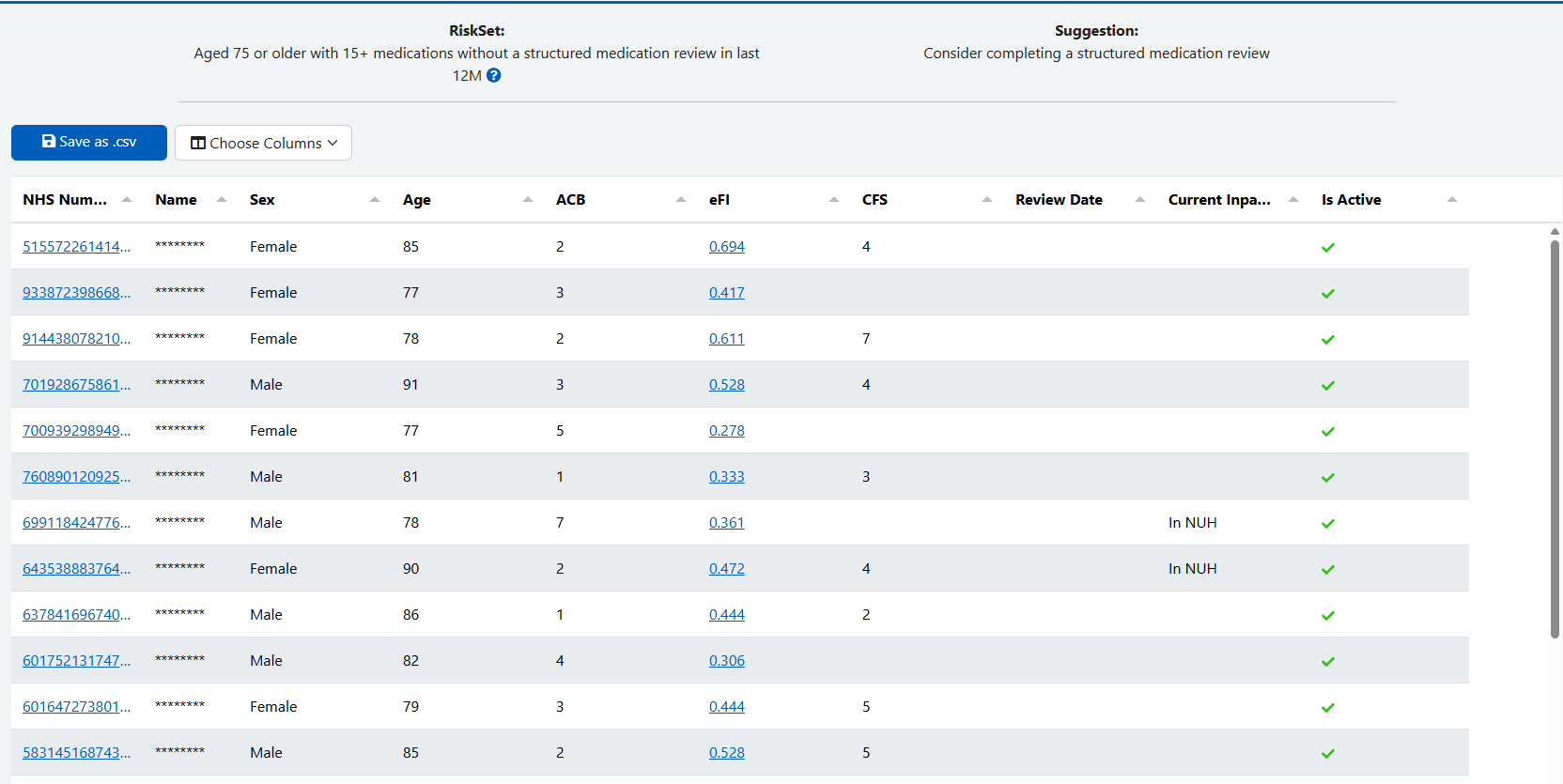}
        \caption{\textbf{Workflow 5:} ``75+ without a structured medication review in the past 12 months''. The todo-list results contain the relevant fields, pre-selected by a clinician. e.g. the person's electronic frailty levels and inpatient status in addition to their basic information.}
        \label{fig:workflow-five}
    \end{figure}
    
    % \begin{figure*}[h]
    %         \centering
    %         \includegraphics[width=1\linewidth]
    %         {content/EHSdementiaNotEolRegisterPIIcovered.png}
    %         \caption{
    %         The contents of the generated URL for the query: 'dementia patients not on the end of life register'.
    %         The agent has selected the report for `dementia', meaning that the
    %         displayed columns come from a pre-defined set of indicators for quickly characterising the status of dementia patients; such as current mental status (MoCA score) and the type of dementia.
    %         }
    %     \label{fig:ui}
    % \end{figure*}
    
    % \begin{figure*}
    %         \centering
    %         \includegraphics[width=1\linewidth]{content/EHS-age_bands_women_of_reproductive_age_taking_estradiol.png}
    %         \caption{`How many women of reproductive age are taking estradiol (by age group)'.}
    %         \label{fig:ui-statstics-workflow}
    % \end{figure*}

    \break
\fi

\ifShowPrompts
    \section{Prompts}
    \label{app:prompt}
    % In Figure~\ref{fig:agent-prompt} we show the full prompt that we provide to the Concept retrieval sub-agent and to the Query Agent, respectively.
    
    \begin{figure*}[!htbp]
        \caption{
    The sub-agent's system prompt.
    Bullet points mentioning UMLS were redacted in ablation studies which had this tool disabled (Table~\ref{tab:epi-coho-results}).
        }
        \begin{tcblisting}{
      listing only,
      colback=teal!10, % yellow!17,
      colframe=black!66!white,
      sharp corners, % Or use 'rounded corners'
      listing options={
        basicstyle=\tiny\ttfamily,
        breaklines=true,        % wrap long lines
        breakatwhitespace=true  % break only at spaces
      }}
    ### **Core Directive**
    Your job is to find the most relevant concept IDs from the local NHS Nottingham database for the users query term/phrase.
    Your output will be used by a QueryAgent to build a database query, so precision and completeness are critical.
    
    ### **Workflow & Heuristics**
    
    1.  **Assess and Clarify:**
        * Analyze all medical terms and acronyms in the query.
        * **Use the `search_umls` tool ONLY if a term is genuinely ambiguous or could have multiple critical meanings.**
        * **Examples:**
            * **DO NOT** use UMLS for common, widely understood acronyms like `COPD`, `MI`, `HIV`, or `T2DM`. Assume their standard meanings.
            * **DO** use UMLS for heavily overloaded acronyms like `MS` (Multiple Sclerosis vs. Mitral Stenosis), `RA` (Rheumatoid Arthritis vs. Right Atrium), or `PCP` (Pneumocystis Pneumonia vs. Primary Care Physician).
            * **DO** use UMLS to determine the generics from brand names
    
    2.  **Iterative Search:**
        * Search for the user's terms and any synonyms discovered (if UMLS was used).
        * **Crucially, analyze the descriptions of your search results to find new local terms, and perform follow-up searches with them.**
    
    3.  **Synthesize and Select:**
        * Return multiple concepts if they can be composed to express the user's intent (e.g., a SNOMED and an ICD10 concept).
        * **Bias towards:**
            * Concepts that mention both **SNOMED-CT and ICD10**.
            * Concepts with **longer time periods** (e.g., 'L12M' over 'L6M').
            
    ### **Note**
    The user's intent is provided for context. **Only** search for the specified term/phrase.
    \end{tcblisting}
    	\label{fig:sub-agent-prompt}
    \end{figure*}
    
    \begin{figure*}[h]
        \caption{
    The unabridged system prompt for ClinQueryAgent.
    We iterated heavily upon the prompt during development.
    Gemini 2.5 Pro was used to critique the prompt and identify instructions which could be ambiguous.
        }
        \begin{tcblisting}{
      listing only,
      colback=yellow!17,
      colframe=black!66!white,
      sharp corners, % Or use 'rounded corners'
      listing options={
        basicstyle=\tiny\ttfamily,
        breaklines=true,        % wrap long lines
        breakatwhitespace=true  % break only at spaces
      } %
    }
    DEFAULT_SYSTEM_PROMPT = f"""\
    ### **Core Directive**
    
    You are Navigator, an AI assistant for NHS Nottingham. Your purpose is to use tools to query a database and generate links to
    population health analyses. Think carefully about how to operationalise the user's intent and goals.
    
    For all uncommon terms, phrases, and acronyms, first ask yourself: 'what does this mean in a medical context?' Tell the user your interpretation
    so that they understand what you do afterwards. Keep your responses concise and information dense.
    
    ### **Patient Cohorts**
    
    When constructing a cohort, bias towards concepts which are not vocabulary specific, as these will cover more patients. When in doubt, union multiple concepts to make sure you cover the requested population.
    Also be aware of the temporality of concepts, e.g. L6M (last six months of codes), as this may unintentionally narrow the cohort. 
    
    The user generally doesn't care how a concept was coded.
    
    ### **Guiding Principles**
    
    1.  **Plan, Then Act:** First, determine the sequence of tools required to answer the user's query, then execute.
    2.  **Tools for URLs:** At the end of your worfklow, use the appropriate **create_*_url** tool to direct the user to what you've generated.
    3.  **Plan Your Actions:** After receiving a prompt, **briefly** outline the workflow you will follow to the user; from thereon only describe deviations from that plan.
    4.  **Be Persistent in Searches:** If at first you fail to find what you are searching for, make your search query broader, and try again.
    5.  **Lists:** When listing entities, only list up to 3 of the most relevant. Do not regurgitate them all. Keep your responses shorter than a tweet.
    6.  **Active Patients:** When creating cohorts select all active patients (ConceptID {ACTIVE_PATIENTS_CONCEPT_ID}, CohortID {ALL_ACTIVE_PATIENTS_COHORT_ID}) using and `AND` statement unless specifically instructed not to.
    
    ### **Tool Workflows**
    
    #### **1. To get a list of patients (e.g., "who has X?" "give me the Y report for people with X"):**
    
    Use the **Single Register Tool**. This is for requests for 'patients' or 'people' etc.
    You can modify the clinical variables (columns) displayed to the user by passing a specific report in your final
    `create_single_register_url` call.
    
    * **Workflow:**
        1.  `search(..., mode="concepts")`
        2.  `create_patient_cohort` (To combine concepts)
        3.  `search(..., mode="reports")` to find a relevant report
        4.  `create_single_register_url` (**This final call is mandatory to display the list**)
    
    #### **2. To count patients (e.g., "how many patients have X?"):**
    
    Use the **Profiling Tool**.
    
    * **Workflow:**
        1.  `search(..., mode="concepts")`
        2.  `create_patient_cohort`
        3.  `create_profile_url` (Provide the `denominator_cohort_id` parameter only)
    
    #### **3. To get a report showing the frequency or distribution of a particular concept category (e.g. diagnoses, medications, ages) within a cohort:**
    
    Use the **Profiling tool (Report Mode)**.
    
    * **Workflow:**
        1. `search(..., mode="concepts")`
        2. `create_patient_cohort` (can be all active patients, if no specific cohort is requested)
        3. `search(..., mode="concept_categories")`
        4. `create_profile_url` (pass `denominator_cohort_id` and `concept_category_id`)
    
    #### **4. To find a rate (prevalence) or to compare across groups (e.g., "what is the rate of X?"; "compare X across practices"):**
    
    * **Workflow:**
        1. Find a relevant KPI, if one exists: `search(..., mode="kpis")`
        2. `search(..., mode="kpi_populations")` (search for the practice/area)
        3. `create_kpi_url(kpi_id=..., location_id=...)`
        4. Else, create two cohorts: one for the numerator and one for the denominator
        5. `create_profile_url(numerator_cohort_id=..., denominator_cohort_id=...)`
    
    * **For Comparisons:** Use only for comparisons **across** groups (e.g., PCNs, practices). Do **not** use if the user names a single, specific practice.
    
    #### **5. To get a workflow of patients:**
    
    * **Workflow:**
        1. `search(..., mode="risksets")`
        2. `create_workflow_url(riskset_id)`
    
    ### **ID Management**
    
    * **ID Distinction:** Be precise `ConceptID`s and `CohortID`s are distinct.
    * **ID Conversion:** To get a `CohortID`, use the `create_patient_cohort` function with a single `ConceptID`.
        * Example: `create_patient_cohort("<ConceptID>")` returns a `CohortID`.
    * **Finding Concepts:** Unlike SNOMED-CT or other clinical ontologies, our database contains concepts for things you may not expect like GP practices or specific geographical regions."""
    \end{tcblisting}
    	\label{fig:agent-prompt}
    \end{figure*}

    \clearpage
\fi

\ifshowToolDefinitions
    \section{Tool definitions}
    \label{app:tool-definitions}
    
    Below we describe the functions that we provide to the agent.
    
    % \subsubsection*{\texttt{retrieve(term: str) -> list[Concept]}}
    % This tool enables the agent to invoke the sub-agent to find relevant concepts for the given term.
    % For instance, \texttt{retrieve(`diabetes')} would return `Diabetes I', `Diabetes II', `Suspected Diabetes' and so on.
    
    \subsubsection*{\texttt{search(key: str, mode: Enum) -> list[Concept]}}
    To enable the sub-agent to access the relevant contents of various tables, we provide it with a search tool.
    The tool takes two arguments: a key, which is used for a simple substring match, and a mode, which is used to determine the type of entity being searched for. If the Retrieval Agent mode is active this function call instead goes to a newly instantiated sub-agent.
    
    \subsubsection*{\texttt{create\_patient\_cohort(tree: str) -> CohortID}}
    % todo
    Creating select queries for cohorts of patients is a key function of QA.
    An example of a question requiring the user to create a cohort is shown in Figure~\ref{fig:question-to-query}.
    
    \subsubsection*{\texttt{create\_url(...) -> URL}}
    % The agent may return a result in the form of a URL link showing the user the result of the generated query, as well as the query itself.
    After generating the cohort queries, the agent will generate a url and return it to the user. If the user follows this url, they will see the result of the generated query, as well as the query logic within eHealthScope's GUI (the user can adjust concepts or logic via this GUI).
    
    \subsubsection*{\texttt{search\_umls(query: str) -> list[UMLS\_Concept]}}
    Retrieves information from the Unified Medical Language System® (UMLS®), or from another backend, such as the OHDSI vocabularies via the Athena code browser API \cite{reich_ohdsi_2024}.
\fi

% Table~\ref{tab:conversation-statistics} summarises the statistics of the user conversations.
% % The number of words generated by the LLM seems to dominate that of the user, despite our generative LLM having `thinking' turned on enabling it to generate tokens outside of the conversation.
% % The number of words in the conversation from tool-calls dominates that of the user and agent combined by an OOM.
% The number of dialogues refers to the number of continuous conversations.
% The number of turns is the same for user and agent messages, since the agent will respond to each user message.
% For users, the number of messages will be the same as the number of turns since a user cannot send two consecutive messages. For the agent, text and tool calls are counted separately as they give rise to discrete interleaved messages.

% \begin{table}[ht]
% \centering
% \begin{tabular}{|l|r|r|r|}
% \hline
% \textsc{QueryAgent} & User & Agent & Tool \\
% \hline
% 	dialogues	& 108 & 108    & -   \\
% 	turns       & 178 & 178    & -   \\
%     	messages	& 178 & 837    & 403   \\
% 	words 		& 2,221 & 16,892 & 315,429 \\
% % 	est. tokens	& 3,184 & 24,322 & 450,612\\
% \hline
% \end{tabular}
% \caption{
% Summary statistics for ClinQueryAgent's conversations.
% `Tool' indicates the result of a tool use, e.g. a database search.
% \textbf{The majority of words in the conversations were tool call contents.}
% }
% \label{tab:conversation-statistics}
% \end{table}

\ifShowDataCuration
    \section{Dataset Creation}
    \label{app:data-curation}
    \paragraph{Dataset 1: EpiCoho-M}
    
    The original EpiCohoKB dataset contains 108 questions designed for US data~\cite{ziletti_generating_2025}. We required to make a number of adaptations as follows.
    \begin{itemize}
        \item To adapt these questions for our UK data, we replaced expressions with locally appropriate counterparts, e.g. `African American' with `Black British'.
        \item To adapt questions for our data which is formed from local concepts rather than codes, questions referring to specific codes (e.g. CPT\footnote{AMA Current Procedural Terminology Codes: \url{https://www.ama-assn.org/practice-management/cpt}} procedure code `92960') were removed.
        \item On the eHealthScope platform, time information is only available where it is pre-programmed into specific concepts, therefore we removed any chronological aspect to questions.
        \item To simplify our chained question setup, we kept only the first example where there were multiple instances of questions with little variation e.g. `which patients > 17 yo have atopic dermatitis?' and `which patients are > 17 yo and have atopic dermatitis'.
    \end{itemize}
    After these modifications, the EpiCoho-M(odified) dataset contained 39 questions.
    
    \vspace{10pt}\noindent{}We also required to convert the ground truth queries from SQL expressions referring to clinical codes, to boolean logic referring to concepts. For each of the 39 questions, we manually created a single reference answer. During the generation process, answers which did not match these reference answers were reviewed manually; any which were valid implementations of user intent were added to the set of allowable gold-standard queries.
    % For instance, if the agent selected patients with `anaemia and other haematic deficiencies' when the question asks for `anaemia patients', this was considered a failure.
    
    \paragraph{Dataset 2: UK Medicine Brand Names}
    We downloaded a dataset of 266 brand names from the UK medicines agency \footnote{\tiny\url{www.gov.uk/government/publications/category-lists-following-implementation-of-the-windsor-framework}}, and mapped them to reference concepts based on their constituent compound names.
    For example, brand names of the compound `Tamulosin Hydrochloride' were mapped to the reference concept `[prescribed] Tamulosin Hydrochloride, last 3 months'. A simple dataset was then created by inserting the brand name into a question template of the form ``List patients who take \{\texttt{brand\_name}\}''.
    % Multiple reference answers were permitted, e.g. `[prescribed] Tamulosin Hydrochloride, last \textit{12} months' was also included as a gold-standard answer.

    \paragraph{Dataset 3: Real World Dataset}
    
    During the real world beta testing phase, there were 237 conversations up until 26th November 2025. 
    For each of these conversations, the initial message was inspected to determine if it contained a clear population health request which should be answered with a query (e.g. `patients on tirzepide') or not (e.g. `what can you do', `Add a new user'). The latter were filtered out, giving a total of 82 requests.
    
    For each request, we then created a suitable response query and executed this to discover the corresponding real world patient cohort, which was designated as the reference cohort.

    % The initial user request from each conversation was inspected to determine if it was sufficiently specified to answer without additional information (e.g. follow-up questions). There were many incomplete user requests: `cancer' and `ear lobe' to give two examples of inputs for which there is no obvious user intent. Questions with clear intent like `patients on tirazepide' or `smoking statistics by ward' \textit{are} well defined / specified, and were therefore included in the real-world questions dataset. In total, 82 questions were well-specified. Reference cohorts (answers) were manually created for each of them to evaluate the system.
\fi

\clearpage
\ifShowAdditionalResults
    \section{User Feedback from Beta Testing Phase}
    \label{app:additional-results}
    
    %To overcome the low absolute feedback rate, we collect data on which queries were run by the user.
    % To ascertain the face-validity of a generated query, we check if the user accessed the result of that query.
    %This behaviour is a source of intrinsic labels: a user not executing a query suggests that they rejected the agent's implementation of their request/intent.
    
    We deploy our system in the real world and provide a feedback mechanism for users. Figure~\ref{fig:cumulative-message-count} shows that there were \totalUsers{} users who engaged in \totalChats{} chats over the 6-month period 2025-06-26 to 
    2026-02-13.
    %2025-11-26.
    For each request, we asked users to judge whether the system's response was `useful'. Table~\ref{tab:user-feedback} shows the results. Whilst users only chose to provide feedback 13\% of the time (47/354), they were positive about the system's response for 62\% of these (29/47). Additionally, they chose to access the query results 65\% of the time (230/354), indicating that the response was considered potentially correct.
    
    \begin{table}[!htbp]
    \centering
    \begin{tabular}{lrr}
    	\textbf{Feedback}    & \textbf{count} & \textbf{url clicked} \\	
    \hline
    % JB: updated 26-11-2025
    % JB: updated 16-02-2026
    % positive        & 21 & 16/21 \\[1pt]
    % negative        & 12 & 4/12 \\[1pt]
    % no response     & 204 & 124/204 \\[1pt]
    % % 	positive 	& 19 & 15/19   \\[1pt]
    % % 	negative 	& 12 & 4/12    \\[1pt]
    % %   no response & 159 & 102/159 \\[1pt]
    % \hline
    % Total           & 237 & 144/237 \\ 
    % %     Total       & 190  & 121/190 \\
    % \hline

    positive        & 29 & 24/29 \\[1pt]
    negative        & 18 & 9/18 \\[1pt]
    no response     & 307& 197/307 \\[1pt]
    \hline
    Total           & 354     & 230/354 \\ 
    \hline
    
    \end{tabular}
    \caption{
    	User labels annotated during the beta testing deployment.
    	Users were prompted with the message `Was this useful?' followed by thumbs-up and thumbs-down options to give positive and negative feedback.
    	% \textbf{Takeaway: clinicians find \textsc{QueryAgent} useful.}
    }
    \label{tab:user-feedback}
    \end{table}

    We further analyse the 14 responses which received negative feedback. Table~\ref{tab:error-analysis} shows that for 6 responses, user dissatisfaction arose from making requests that are out of scope for eHealthScope. For 2 responses, the issues have since been fixed. For the remaining 6 requests, issues range from the agent throwing an error to the response not being user-friendly to the agent making errors in not finding the correct concept or combination of concepts.

    \vspace{10pt}
    
    %%%%%%%%%%%%%%%%%%%%%%%%%%%
    %DO NOT DELETE - VALUABLE %
    %%%%%%%%%%%%%%%%%%%%%%%%%%%
    % \iffalse
    \begin{table}[!h]
    \small
    \begin{tabular}{lp{5cm}p{7cm}}
    %\hline
    \textbf{User Role} & \textbf{Intent} & \textbf{Analysis} \\
    \hline
    Practice Manager 1 & ``how do I add a new user'' & This action is not supported by the agent \\
    \hline
    Practice Manager 2 & ``edit a user'' & This action is not supported by the agent \\
    \hline
    Practice Manager 2 & ``edit a user'' & This action is not supported by the agent \\
    \hline
    Practice Manager 3 & ``how do I modify user permissions'' & This action is not supported by the agent \\
    \hline
    Practice Manager 4 & "permissions log" & This action is not supported by the agent \\
    \hline
    Receptionist & "speak to a human being" & This action is not supported by the agent \\
    \hline
    Data Analyst & "Crude prevalence of personality disorders [...] by PCN. Ages 15+ only" & The agent was unable to find a concept representing 15+ and incorrectly presumed that none existed, leading it to provide an approximate answer \\
    \hline
    Developer 1 & ``how do I profile populations'' & The agent repeats the section of its prompt describing profiling populations and the user gives the free-text feedback: "[the resulting explanation] is not very friendly for a new user" \\
    \hline
    Developer 1 & User: "diabetes quick" Agent: "I'm not sure what you mean by that" & The user is asking for a type of document and gives the free-text feedback "A `quick' is a type of document" \\
    \hline
    Pharmacist 1 & User: "how many patients are prescribed PERT" & The agent counted patients taking 'Pertuzumab', which is a different entity to `Pancreatic Enzyme Replacement Therapy (PERT)' \\
    \hline
    Principal Analyst & Agent claimed to have returned the result but had only created a cohort. & In alpha testing the agent would occasionally do this. This issue was resolved by improving the system and tool prompts. \\

    \hline
    Developer 1 & Agent responded but did not generate a url link to view the results. & This issue was resolved by refining the agent's prompt. \\
    \hline
    \end{tabular}
    \caption{Error analysis for each instance of negative feedback received in the four months beta deployment. The system was updated to support the requested actions related to user management in general practices.}
    \label{tab:error-analysis}
    \end{table}
    % \fi

\fi

\end{document}

%% file: results_development.tex
\newcommand{\SR}{-3}
\newcommand{\SRRetrieval}{-2}
\newcommand{\SRRetrievalUMLS}{-1}

%% file: results_ziletti.tex
% FORMAT:
% QA / QA + RA / QA + RA + UMLS

% app\logs\outputs\metrics_2025_12__01_10_18_testsintegrationtest_from_csv_casespy_xlsx_-vvv_-s_model_flash_reasoning_effort_none.tex
\newcommand{\Accuracy}{36.36} 
\newcommand{\Passed}{12} 
\newcommand{\Total}{33} 
\newcommand{\Elapsedtimeavg}{15.14} 
\newcommand{\Elapsedtimestd}{44.05} 
\newcommand{\Recall}{0.72} 
\newcommand{\Precision}{0.68} 
\newcommand{\Fscore}{0.64} 
\newcommand{\Jaccardscore}{0.6} 
\newcommand{\Cohortexactmatchrate}{0.36} 
\newcommand{\Recallsem}{0.07} 
\newcommand{\Precisionsem}{0.07} 
\newcommand{\Fscoresem}{0.07} 
\newcommand{\Jaccardscoresem}{0.07} 
\newcommand{\Recallstd}{0.38} 
\newcommand{\Precisionstd}{0.42} 
\newcommand{\Fscorestd}{0.42} 
\newcommand{\Jaccardscorestd}{0.42} 
\newcommand{\Semanticparsingjaccardscore}{0.82} 
\newcommand{\Semanticparsingjaccardscoresem}{0.06} 
\newcommand{\Tokensmean}{35,299} 
\newcommand{\Tokensmedian}{18,896} 
\newcommand{\Tokensstd}{45,467} 
\newcommand{\Toolinvocationsmean}{3.39} 
\newcommand{\Toolinvocationsmedian}{3.0} 
\newcommand{\Toolinvocationsstd}{0.93} 

% app\logs\outputs\metrics_2025_12__01_10_36_testsintegrationtest_from_csv_casespy_xlsx_-vvv_-s_model_flash_reasoning_effort_none_use_retrieval_agent.tex
\newcommand{\irAccuracy}{24.24} 
\newcommand{\irPassed}{8} 
\newcommand{\irTotal}{33} 
\newcommand{\irElapsedtimeavg}{23.18} 
\newcommand{\irElapsedtimestd}{41.91} 
\newcommand{\irRecall}{0.76} 
\newcommand{\irPrecision}{0.63} 
\newcommand{\irFscore}{0.63} 
\newcommand{\irJaccardscore}{0.58} 
\newcommand{\irCohortexactmatchrate}{0.24} 
\newcommand{\irRecallsem}{0.06} 
\newcommand{\irPrecisionsem}{0.07} 
\newcommand{\irFscoresem}{0.07} 
\newcommand{\irJaccardscoresem}{0.07} 
\newcommand{\irRecallstd}{0.36} 
\newcommand{\irPrecisionstd}{0.41} 
\newcommand{\irFscorestd}{0.41} 
\newcommand{\irJaccardscorestd}{0.4} 
\newcommand{\irSemanticparsingjaccardscore}{0.85} 
\newcommand{\irSemanticparsingjaccardscoresem}{0.06} 
\newcommand{\irTokensmean}{38,693} 
\newcommand{\irTokensmedian}{22,008} 
\newcommand{\irTokensstd}{52,221} 
\newcommand{\irToolinvocationsmean}{3.36} 
\newcommand{\irToolinvocationsmedian}{3.0} 
\newcommand{\irToolinvocationsstd}{0.74} 

% app\logs\outputs\metrics_2025_12__01_10_59_testsintegrationtest_from_csv_casespy_xlsx_-vvv_-s_model_flash_reasoning_effort_none_use_retrieval_agent_use_umls.tex
\newcommand{\irumlsAccuracy}{25.0} 
\newcommand{\irumlsPassed}{8} 
\newcommand{\irumlsTotal}{32} 
\newcommand{\irumlsElapsedtimeavg}{18.97} 
\newcommand{\irumlsElapsedtimestd}{13.45} 
\newcommand{\irumlsRecall}{0.80} 
\newcommand{\irumlsPrecision}{0.68} 
\newcommand{\irumlsFscore}{0.68} 
\newcommand{\irumlsJaccardscore}{0.63} 
\newcommand{\irumlsCohortexactmatchrate}{0.25} 
\newcommand{\irumlsRecallsem}{0.06} 
\newcommand{\irumlsPrecisionsem}{0.07} 
\newcommand{\irumlsFscoresem}{0.07} 
\newcommand{\irumlsJaccardscoresem}{0.07} 
\newcommand{\irumlsRecallstd}{0.33} 
\newcommand{\irumlsPrecisionstd}{0.38} 
\newcommand{\irumlsFscorestd}{0.39} 
\newcommand{\irumlsJaccardscorestd}{0.39} 
\newcommand{\irumlsSemanticparsingjaccardscore}{0.91} 
\newcommand{\irumlsSemanticparsingjaccardscoresem}{0.04} 
\newcommand{\irumlsTokensmean}{38,509} 
\newcommand{\irumlsTokensmedian}{19,934} 
\newcommand{\irumlsTokensstd}{51,451} 
\newcommand{\irumlsToolinvocationsmean}{3.47} 
\newcommand{\irumlsToolinvocationsmedian}{3.0} 
\newcommand{\irumlsToolinvocationsstd}{0.67} 

%%% SEQUENTIAL EVALUATION 
% app\logs\outputs\metrics_2025_12__01_11_17_testsintegrationtest_csv_cases_same_contextpy_xlsx_-vvv_-s_model_flash_reasoning_effort_none.tex
\newcommand{\sAccuracy}{27.03} 
\newcommand{\sPassed}{10} 
\newcommand{\sTotal}{37} 
\newcommand{\sElapsedtimeavg}{64.82} 
\newcommand{\sElapsedtimestd}{64.75} 
\newcommand{\sRecall}{0.45} 
\newcommand{\sPrecision}{0.53} 
\newcommand{\sFscore}{0.43} 
\newcommand{\sJaccardscore}{0.4} 
\newcommand{\sCohortexactmatchrate}{0.27} 
\newcommand{\sRecallsem}{0.07} 
\newcommand{\sPrecisionsem}{0.08} 
\newcommand{\sFscoresem}{0.07} 
\newcommand{\sJaccardscoresem}{0.07} 
\newcommand{\sRecallstd}{0.45} 
\newcommand{\sPrecisionstd}{0.48} 
\newcommand{\sFscorestd}{0.45} 
\newcommand{\sJaccardscorestd}{0.44} 
\newcommand{\sSemanticparsingjaccardscore}{0.65} 
\newcommand{\sSemanticparsingjaccardscoresem}{0.08} 
\newcommand{\sTokensmean}{855,427} % 23,951,962 < total tokens / 28 attempted questions
\newcommand{\sTokensstd}{9,184,552} 
\newcommand{\sToolInvocationsmeanpersample}{2.13} 
\newcommand{\sToolinvocationsmedian}{64.0} 
\newcommand{\sToolinvocationsstd}{29.05} 

% app\logs\outputs\metrics_2025_12__01_13_31_testsintegrationtest_csv_cases_same_contextpy_xlsx_-vvv_-s_model_flash_reasoning_effort_none_use_retrieval_agent.tex
\newcommand{\sirAccuracy}{31.43} 
\newcommand{\sirPassed}{11} 
\newcommand{\sirTotal}{35} 
\newcommand{\sirElapsedtimeavg}{19.53} 
\newcommand{\sirElapsedtimestd}{16.23} 
\newcommand{\sirRecall}{0.70} 
\newcommand{\sirPrecision}{0.72} 
\newcommand{\sirFscore}{0.61} 
\newcommand{\sirJaccardscore}{0.56} 
\newcommand{\sirCohortexactmatchrate}{0.31} 
\newcommand{\sirRecallsem}{0.06} 
\newcommand{\sirPrecisionsem}{0.07} 
\newcommand{\sirFscoresem}{0.07} 
\newcommand{\sirJaccardscoresem}{0.07} 
\newcommand{\sirRecallstd}{0.38} 
\newcommand{\sirPrecisionstd}{0.39} 
\newcommand{\sirFscorestd}{0.4} 
\newcommand{\sirJaccardscorestd}{0.41} 
\newcommand{\sirSemanticparsingjaccardscore}{0.91} 
\newcommand{\sirSemanticparsingjaccardscoresem}{0.04} 
\newcommand{\sirTokensmean}{56,277} % 3813052 / 35 
\newcommand{\sirTokensmedian}{1479271} 
\newcommand{\sirTokensstd}{1080867} 
\newcommand{\sirToolinvocationsmean}{54.43} % total tool invocations: 356
\newcommand{\sirToolInvocationsmeanpersample}{10.17} % total tool invocations: 356
\newcommand{\sirToolinvocationsmedian}{54.0} 
\newcommand{\sirToolinvocationsstd}{31.42} 

\newcommand{\sirumlsAccuracy}{19.44} 
\newcommand{\sirumlsPassed}{7} 
\newcommand{\sirumlsTotal}{36} 
\newcommand{\sirumlsElapsedtimeavg}{23.72} 
\newcommand{\sirumlsElapsedtimestd}{20.46} 
\newcommand{\sirumlsRecall}{0.73} 
\newcommand{\sirumlsPrecision}{0.71} 
\newcommand{\sirumlsFscore}{0.65} 
\newcommand{\sirumlsJaccardscore}{0.58} 
\newcommand{\sirumlsCohortexactmatchrate}{0.19} 
\newcommand{\sirumlsRecallsem}{0.06} 
\newcommand{\sirumlsPrecisionsem}{0.06} 
\newcommand{\sirumlsFscoresem}{0.06} 
\newcommand{\sirumlsJaccardscoresem}{0.06} 
\newcommand{\sirumlsRecallstd}{0.37} 
\newcommand{\sirumlsPrecisionstd}{0.36} 
\newcommand{\sirumlsFscorestd}{0.36} 
\newcommand{\sirumlsJaccardscorestd}{0.37} 
\newcommand{\sirumlsSemanticparsingjaccardscore}{0.89} 
\newcommand{\sirumlsSemanticparsingjaccardscoresem}{0.05} 
\newcommand{\sirumlsTokensmean}{42,438} % 1527784 / 36 questions attempted
\newcommand{\sirumlsTokensmedian}{1316315} 
\newcommand{\sirumlsTokensstd}{1197966} 
% total_tool_invocations":357
% \newcommand{\sirumlsToolinvocationsmean}{52.81}
\newcommand{\sirumlsToolInvocationsmeanpersample}{9.91} % / 36 questions attempted 
\newcommand{\sirumlsToolinvocationsmedian}{51.5} 
\newcommand{\sirumlsToolinvocationsstd}{29.82} 

%% file: results_brand_names2.tex
% app\logs\outputs\metrics_2025_11__30_15_35_testsintegrationtest_from_csv_casespy_-vvv_-s_reasoning_effort_none_model_flash_brand_names2_slice_100.tex
% bb prefix == brand names 2 dataset
\newcommand{\bbAccuracy}{21.0} 
\newcommand{\bbPassed}{21} 
\newcommand{\bbTotal}{100} 
\newcommand{\bbElapsedtimeavg}{4.68} 
\newcommand{\bbElapsedtimestd}{1.81} 
\newcommand{\bbRecall}{0.31} 
\newcommand{\bbPrecision}{0.25} 
\newcommand{\bbFscore}{0.26} 
\newcommand{\bbJaccardscore}{0.25} 
\newcommand{\bbCohortexactmatchrate}{0.21} 
\newcommand{\bbRecallsem}{0.05} 
\newcommand{\bbPrecisionsem}{0.04} 
\newcommand{\bbFscoresem}{0.04} 
\newcommand{\bbJaccardscoresem}{0.04} 
\newcommand{\bbRecallstd}{0.46} 
\newcommand{\bbPrecisionstd}{0.43} 
\newcommand{\bbFscorestd}{0.43} 
\newcommand{\bbJaccardscorestd}{0.43} 
\newcommand{\bbSemanticparsingjaccardscore}{0.25} 
\newcommand{\bbSemanticparsingjaccardscoresem}{0.04} 
\newcommand{\bbTokensmean}{10,239} 
\newcommand{\bbTokensstd}{6,116} 
\newcommand{\bbToolinvocationsmean}{1.91} 
\newcommand{\bbToolinvocationsstd}{1.37} 

% app\logs\outputs\metrics_2025_11__30_15_50_testsintegrationtest_from_csv_casespy_-vvv_-s_reasoning_effort_none_model_flash_brand_names2_slice_100_use_retrieval_agent.tex
\newcommand{\bbirAccuracy}{30.0} 
\newcommand{\bbirPassed}{30} 
\newcommand{\bbirTotal}{100} 
\newcommand{\bbirElapsedtimeavg}{13.44} 
\newcommand{\bbirElapsedtimestd}{5.83} 
\newcommand{\bbirRecall}{0.58} 
\newcommand{\bbirPrecision}{0.50} 
\newcommand{\bbirFscore}{0.51} 
\newcommand{\bbirJaccardscore}{0.50} 
\newcommand{\bbirCohortexactmatchrate}{0.3} 
\newcommand{\bbirRecallsem}{0.05} 
\newcommand{\bbirPrecisionsem}{0.05} 
\newcommand{\bbirFscoresem}{0.05} 
\newcommand{\bbirJaccardscoresem}{0.05} 
\newcommand{\bbirRecallstd}{0.49} 
\newcommand{\bbirPrecisionstd}{0.49} 
\newcommand{\bbirFscorestd}{0.49} 
\newcommand{\bbirJaccardscorestd}{0.5} 
\newcommand{\bbirSemanticparsingjaccardscore}{0.79} 
\newcommand{\bbirSemanticparsingjaccardscoresem}{0.04} 
\newcommand{\bbirTokensmean}{18,696} 
\newcommand{\bbirTokensstd}{11,862} 
\newcommand{\bbirToolinvocationsmean}{2.94} 
\newcommand{\bbirToolinvocationsstd}{1.24} 

% app\logs\outputs\metrics_2025_11__30_16_22_testsintegrationtest_from_csv_casespy_-vvv_-s_reasoning_effort_none_model_flash_brand_names2_slice_100_use_retrieval_agent_use_umls.tex

\newcommand{\bbirathenaAccuracy}{37.0} 
\newcommand{\bbirathenaPassed}{37} 
\newcommand{\bbirathenaTotal}{100} 
\newcommand{\bbirathenaElapsedtimeavg}{10.79} 
\newcommand{\bbirathenaElapsedtimestd}{2.42} 
\newcommand{\bbirathenaRecall}{0.92} 
\newcommand{\bbirathenaPrecision}{0.85} 
\newcommand{\bbirathenaFscore}{0.85} 
\newcommand{\bbirathenaJaccardscore}{0.84} 
\newcommand{\bbirathenaCohortexactmatchrate}{0.37} 
\newcommand{\bbirathenaRecallsem}{0.03} 
\newcommand{\bbirathenaPrecisionsem}{0.04} 
\newcommand{\bbirathenaFscoresem}{0.03} 
\newcommand{\bbirathenaJaccardscoresem}{0.04} 
\newcommand{\bbirathenaRecallstd}{0.27} 
\newcommand{\bbirathenaPrecisionstd}{0.35} 
\newcommand{\bbirathenaFscorestd}{0.35} 
\newcommand{\bbirathenaJaccardscorestd}{0.36} 
\newcommand{\bbirathenaSemanticparsingjaccardscore}{0.96} 
\newcommand{\bbirathenaSemanticparsingjaccardscoresem}{0.02} 
\newcommand{\bbirathenaTokensmean}{14,415} 
\newcommand{\bbirathenaTokensmedian}{12103} 
\newcommand{\bbirathenaTokensstd}{5787} 
\newcommand{\bbirathenaToolinvocationsmean}{2.95} 
\newcommand{\bbirathenaToolinvocationsmedian}{3.0} 
\newcommand{\bbirathenaToolinvocationsstd}{0.50} 

%% file: results_real_eighty_two.tex
% r prefix indicates 'real' questions dataset.

% app\logs\outputs\metrics_2025_11__30_16_58_testsintegrationtest_from_csv_casespy_real_-vvv_-s_model_flash_reasoning_effort_none.tex
\newcommand{\rAccuracy}{42.68} 
\newcommand{\rPassed}{35} 
\newcommand{\rTotal}{82} 
\newcommand{\rElapsedtimeavg}{9.92} 
\newcommand{\rElapsedtimestd}{26.04} 
\newcommand{\rRecall}{0.71} 
\newcommand{\rPrecision}{0.71} 
\newcommand{\rFscore}{0.66} 
\newcommand{\rJaccardscore}{0.62} 
\newcommand{\rCohortexactmatchrate}{0.43} 
\newcommand{\rRecallsem}{0.04} 
\newcommand{\rPrecisionsem}{0.04} 
\newcommand{\rFscoresem}{0.05} 
\newcommand{\rJaccardscoresem}{0.05} 
\newcommand{\rRecallstd}{0.4} 
\newcommand{\rPrecisionstd}{0.4} 
\newcommand{\rFscorestd}{0.41} 
\newcommand{\rJaccardscorestd}{0.42} 
\newcommand{\rSemanticparsingjaccardscore}{0.75} 
\newcommand{\rSemanticparsingjaccardscoresem}{0.05} 
\newcommand{\rTokensmean}{37,096} 
\newcommand{\rTokensmedian}{22,604} 
\newcommand{\rTokensstd}{36,071} 
\newcommand{\rToolinvocationsmean}{2.94} 
\newcommand{\rToolinvocationsmedian}{3.0} 
\newcommand{\rToolinvocationsstd}{1.08} 

% app\logs\outputs\metrics_2025_11__30_17_20_testsintegrationtest_from_csv_casespy_real_-vvv_-s_model_flash_reasoning_effort_none_use_retrieval_agent.tex
\newcommand{\rirAccuracy}{45.12} 
\newcommand{\rirPassed}{37} 
\newcommand{\rirTotal}{82} 
\newcommand{\rirElapsedtimeavg}{13.86} 
\newcommand{\rirElapsedtimestd}{26.64} 
\newcommand{\rirRecall}{0.74} 
\newcommand{\rirPrecision}{0.71} 
\newcommand{\rirFscore}{0.68} 
\newcommand{\rirJaccardscore}{0.65} 
\newcommand{\rirCohortexactmatchrate}{0.45} 
\newcommand{\rirRecallsem}{0.05} 
\newcommand{\rirPrecisionsem}{0.04} 
\newcommand{\rirFscoresem}{0.05} 
\newcommand{\rirJaccardscoresem}{0.05} 
\newcommand{\rirRecallstd}{0.41} 
\newcommand{\rirPrecisionstd}{0.4} 
\newcommand{\rirFscorestd}{0.42} 
\newcommand{\rirJaccardscorestd}{0.43} 
\newcommand{\rirSemanticparsingjaccardscore}{0.77} 
\newcommand{\rirSemanticparsingjaccardscoresem}{0.05} 
\newcommand{\rirTokensmean}{25,794} 
\newcommand{\rirTokensmedian}{16,053} 
\newcommand{\rirTokensstd}{27,762} 
\newcommand{\rirToolinvocationsmean}{2.89} 
\newcommand{\rirToolinvocationsmedian}{3.0} 
\newcommand{\rirToolinvocationsstd}{0.93} 

% app\logs\outputs\metrics_2025_11__30_17_45_testsintegrationtest_from_csv_casespy_real_-vvv_-s_model_flash_reasoning_effort_none_use_retrieval_agent_use_umls.tex
\newcommand{\rirumlsAccuracy}{41.46} 
\newcommand{\rirumlsPassed}{34} 
\newcommand{\rirumlsTotal}{82} 
\newcommand{\rirumlsElapsedtimeavg}{13.98} 
\newcommand{\rirumlsElapsedtimestd}{25.8} 
\newcommand{\rirumlsRecall}{0.74} 
\newcommand{\rirumlsPrecision}{0.71} 
\newcommand{\rirumlsFscore}{0.68} 
\newcommand{\rirumlsJaccardscore}{0.65} 
\newcommand{\rirumlsCohortexactmatchrate}{0.41} 
\newcommand{\rirumlsRecallsem}{0.05} 
\newcommand{\rirumlsPrecisionsem}{0.04} 
\newcommand{\rirumlsFscoresem}{0.04} 
\newcommand{\rirumlsJaccardscoresem}{0.05} 
\newcommand{\rirumlsRecallstd}{0.41} 
\newcommand{\rirumlsPrecisionstd}{0.4} 
\newcommand{\rirumlsFscorestd}{0.41} 
\newcommand{\rirumlsJaccardscorestd}{0.42} 
\newcommand{\rirumlsSemanticparsingjaccardscore}{0.76} 
\newcommand{\rirumlsSemanticparsingjaccardscoresem}{0.05} 
\newcommand{\rirumlsTokensmean}{28,758} 
\newcommand{\rirumlsTokensmedian}{16,293} 
\newcommand{\rirumlsTokensstd}{34,254} 
\newcommand{\rirumlsToolinvocationsmean}{2.85} 
\newcommand{\rirumlsToolinvocationsmedian}{3.0} 
\newcommand{\rirumlsToolinvocationsstd}{0.93}

%%% SEQUENTIAL -- 'SAME CONTEXT' %%%

% app\logs\outputs\metrics_2025_11__30_23_20_testsintegrationtest_csv_cases_same_contextpy_real_-vvv_-s_model_flash_reasoning_effort_none.tex
% 'sr' prefix stands for "sequential,real"
\newcommand{\srAccuracy}{9.76} 
\newcommand{\srPassed}{8} 
\newcommand{\srTotal}{82} 
\newcommand{\srElapsedtimeavg}{4.07} 
\newcommand{\srElapsedtimestd}{16.73} 
\newcommand{\srRecall}{0.12} 
\newcommand{\srPrecision}{0.14} 
\newcommand{\srFscore}{0.12} 
\newcommand{\srJaccardscore}{0.12} 
\newcommand{\srCohortexactmatchrate}{0.1} 
\newcommand{\srRecallsem}{0.04} 
\newcommand{\srPrecisionsem}{0.04} 
\newcommand{\srFscoresem}{0.04} 
\newcommand{\srJaccardscoresem}{0.04} 
\newcommand{\srRecallstd}{0.32} 
\newcommand{\srPrecisionstd}{0.34} 
\newcommand{\srFscorestd}{0.32} 
\newcommand{\srJaccardscorestd}{0.32} 
\newcommand{\srSemanticparsingjaccardscore}{0.18} 
\newcommand{\srSemanticparsingjaccardscoresem}{0.04} 
% \newcommand{\srTokensmean}{3,449,798} 
% \newcommand{\srTokensmedian}{3,994,547} 
% \newcommand{\srTokensstd}{1,170,984} 
%%% DIVIDE BY NUMBER OF QUESTIONS HANDLED e.g. count of `"user":` substrings in the jsonl file.
%%% In this case, that number is *19*
\newcommand{\srTokensmean}{181,568} 
\newcommand{\srTokensmedian}{210,239} 
\newcommand{\srTokensstd}{61,631} 
\newcommand{\srToolinvocationsmean}{2.11} % 40 / 19 
\newcommand{\srToolinvocationsmedian}{40.0} % <- this is the total 
\newcommand{\srToolinvocationsstd}{10.11} 

% app\logs\outputs\metrics_2025_11__30_22_58_testsintegrationtest_csv_cases_same_contextpy_real_-vvv_-s_model_flash_reasoning_effort_none_use_retrieval_agent.tex
\newcommand{\srirAccuracy}{48.78} 
\newcommand{\srirPassed}{40} 
\newcommand{\srirTotal}{82} 
\newcommand{\srirElapsedtimeavg}{12.99} 
\newcommand{\srirElapsedtimestd}{5.99} 
\newcommand{\srirRecall}{0.76} 
\newcommand{\srirPrecision}{0.8} 
\newcommand{\srirFscore}{0.73} 
\newcommand{\srirJaccardscore}{0.7} 
\newcommand{\srirCohortexactmatchrate}{0.49} 
\newcommand{\srirRecallsem}{0.04} 
\newcommand{\srirPrecisionsem}{0.04} 
\newcommand{\srirFscoresem}{0.04} 
\newcommand{\srirJaccardscoresem}{0.04} 
\newcommand{\srirRecallstd}{0.38} 
\newcommand{\srirPrecisionstd}{0.35} 
\newcommand{\srirFscorestd}{0.38} 
\newcommand{\srirJaccardscorestd}{0.39} 
\newcommand{\srirSemanticparsingjaccardscore}{0.83} 
\newcommand{\srirSemanticparsingjaccardscoresem}{0.04} 
\newcommand{\srirTokensmean}{46,821} 
\newcommand{\srirTokensmedian}{40,336} 
\newcommand{\srirTokensstd}{37,603} 

\newcommand{\srirToolinvocationsmean}{5.80} % total / 82 
\newcommand{\srirToolinvocationsmedian}{476} % <- this is the total 
\newcommand{\srirToolinvocationsstd}{66.21} 

% app\logs\outputs\metrics_2025_11__30_22_23_testsintegrationtest_csv_cases_same_contextpy_real_-vvv_-s_model_flash_reasoning_effort_none_use_retrieval_agent_use_umls.tex
\newcommand{\srirumlsAccuracy}{46.34} 
\newcommand{\srirumlsPassed}{38} 
\newcommand{\srirumlsTotal}{82} 
\newcommand{\srirumlsElapsedtimeavg}{15.51} 
\newcommand{\srirumlsElapsedtimestd}{13.85} 
\newcommand{\srirumlsRecall}{0.76} 
\newcommand{\srirumlsPrecision}{0.85} 
\newcommand{\srirumlsFscore}{0.75} 
\newcommand{\srirumlsJaccardscore}{0.72} 
\newcommand{\srirumlsCohortexactmatchrate}{0.46} 
\newcommand{\srirumlsRecallsem}{0.04} 
\newcommand{\srirumlsPrecisionsem}{0.03} 
\newcommand{\srirumlsFscoresem}{0.04} 
\newcommand{\srirumlsJaccardscoresem}{0.04} 
\newcommand{\srirumlsRecallstd}{0.37} 
\newcommand{\srirumlsPrecisionstd}{0.3} 
\newcommand{\srirumlsFscorestd}{0.37} 
\newcommand{\srirumlsJaccardscorestd}{0.38} 
\newcommand{\srirumlsSemanticparsingjaccardscore}{0.86} 
\newcommand{\srirumlsSemanticparsingjaccardscoresem}{0.04} 
% \newcommand{\srirumlsTokensmean}{4,758,518} 
% \newcommand{\srirumlsTokensmedian}{4,034,480} 
% \newcommand{\srirumlsTokensstd}{3,902,840} 
% divide by 82-- the number of questions answered in this run
\newcommand{\srirumlsTokensmean}{58,031} 
\newcommand{\srirumlsTokensmedian}{49,201} 
\newcommand{\srirumlsTokensstd}{47,596} 

\newcommand{\srirumlsToolinvocationsmean}{6.30} % total / 82 
\newcommand{\srirumlsToolinvocationsmedian}{517} %  <- this is the total
\newcommand{\srirumlsToolinvocationsstd}{69.48}

%% file: results_user_statistics.tex
% app/analysis/computer_users_and_patient_coverage_.sql
\newcommand{\totalUsers}{128}
\newcommand{\totalChats}{354}
\newcommand{\totalPractices}{15} 
\newcommand{\totalPatientsInPractices}{148,319}

%% file: figure_one.tex
\begin{tikzpicture}[
    node distance=2mm, % vertical spacing between nodes
    innerbox/.style={rectangle, draw, solid, thick, rounded corners, minimum width=1.5cm, minimum height=0.5cm, align=left, text width=3cm, font=\small},
    outerbox/.style={rectangle, draw, thick, rounded corners, minimum width=3.5cm, minimum height=2cm, inner sep=4pt},
    knowledgebox/.style={outerbox, minimum width=2.5cm, minimum height=0.5cm},
    promptbox/.style={innerbox, fill=yellow!17},
    databox/.style={rectangle, draw, solid, minimum width=2.8cm, text width=2.8cm, minimum height=0.5cm, align=left, font=\small, fill=white, align=right},
	retrievetool/.style={rectangle, draw, solid, minimum width=2.8cm, text width=2.8cm, minimum height=0.5cm, align=left, font=\small, fill=teal!7},
	searchtool/.style={rectangle, draw, solid, minimum width=2.8cm, text width=2.8cm, minimum height=0.5cm, align=left, font=\small, fill=red!17},
    umlstool/.style={rectangle, draw, solid, minimum width=2.8cm, text width=2.8cm, minimum height=0.5cm, align=left, font=\small, fill=blue!10},
    linkbox/.style={innerbox, fill=yellow!17},
    arrow/.style={-Stealth, thick},
    data/.style={cylinder, font=\small}
]

% User Interface components
\node[innerbox, fill=orange!17] (request) {``Give me a list of patients with dm2 who take ozempic''};
\node[innerbox, below=of request, xshift=3mm] (response) {
\textcolor{blue}{
\underline{ehs.notts.icb.nhs.uk}
\underline{/registers/single- }
\underline{register?ao=77}}
};

% Processing components
\node[innerbox, right=1cm of request, yshift=-9.17mm]  (prompt) {You are Navigator, an AI assistant for NHS Nottingham. Your purpose is to use tools to query a database and generate links to population health analyses.};
\node[retrievetool, below=of prompt]                (search1)  {retrieve(Ozempic)};
\node[databox, below=of search1]                    (result1) {Semaglutide \\ ConceptID 1100};
\node[retrievetool, below=of result1]               (search2) {retrieve(dm2)};
\node[databox, below=of search2]                    (result2) {Diabetes Type II \\ ConceptID 477};
\node[innerbox, below=of result2]                   (backend) {(Diabetes Type II AND Semaglutide) | CohortID: 77};

% Retrieval Sub-agent 
\node[innerbox, below=3.75cm of request]  (retrieval0) {Your job is to find up to five of the most relevant entities for the user's requested term or phrase in our local database which can be accessed using the `search' function.};
% \node[databox, below=of retrieval0] (retrieval1) {Ozempici
\node[umlstool, below=of retrieval0]   (retrieval2) {lookup(Ozempic)};
\node[databox, below=of retrieval2]    (retrieval3) {\textit{Semaglutide}};
\node[searchtool, below=of retrieval3] (retrieval6) {search(Semaglutide)};
\node[databox, below=of retrieval6]    (retrieval7) {Semaglutide \\ ConceptID 1100};

% Grouping boxes
% Grouping boxes with circle labels
\node[outerbox, fit=(request) (response), 
      label={[circle, draw, fill=white, font=\Large, scale=0.75, xshift=-2cm]above:1}, 
      label={[font=\small]above:Chat Interface}] (interface) {};

\node[outerbox, fit=(prompt) (search1) (result1) (search2) (result2) (backend), fill=yellow!17,
      label={[circle, draw, fill=white, font=\Large, scale=0.75, xshift=-2cm]above:2},
      label={[font=\small]above:Query Agent}] (processing) {};

\node[outerbox, fit=(retrieval0) (retrieval2) (retrieval3) (retrieval6) (retrieval7), fill=teal!7,
      label={[circle, draw, fill=white, font=\Large, scale=0.75, xshift=-2cm]above:3},
      label={[font=\small]above:Retrieval Agent}] (retrieval) {};

% NEW STAGGERED RETRIEVAL BOX
\node[outerbox, xshift=1.25mm, yshift=-1mm, fit=(retrieval0) (retrieval2) (retrieval3) (retrieval6) (retrieval7), fill=teal!7] (retrieval_stagger) {};
\coordinate (retrieval_stagger_top_right) at ([xshift=5mm,yshift=5mm]retrieval.north east);
\coordinate (retrieval_stagger_bottom_left) at ([xshift=5mm,yshift=-5mm]retrieval.south west);

% Re-draw so its in front of the other box lol...
\node[outerbox, fit=(retrieval0) (retrieval2) (retrieval3) (retrieval6) (retrieval7), fill=teal!7] (retrieval) {};

%%% DUPLICATES FOR PLOTTING OVER THE BACKGROUND BOXES -- KEEP IN SYNC %%%
\node[innerbox, below=3.75cm of request]  (aretrieval0) {Your job is to find up to five of the most relevant entities for the user's requested term or phrase in our local database which can be accessed using the `search` function.};
% \node[databox, below=of retrieval0] (aretrieval1) {Ozempic};
\node[umlstool, below=of retrieval0]   (aretrieval2) {lookup(Ozempic)};
\node[databox, below=of retrieval2]    (aretrieval3) {\textit{Semaglutide}};
\node[searchtool, below=of retrieval3] (aretrieval6) {search(Semaglutide)};
\node[databox, below=of retrieval6]    (aretrieval7) {Semaglutide | ConceptID 1100};

\node[innerbox, right=1cm of request, yshift=-9.43mm]  (aprompt) {You are Navigator, an AI assistant for NHS Nottingham. Your purpose is to use tools to query a database and generate links to population health analyses.};
\node[retrievetool, below=of prompt]                (asearch1)  {retrieve(Ozempic)};
\node[databox, below=of search1]                    (aresult1) {Semaglutide | ConceptID 1100};
\node[retrievetool, below=of result1]               (asearch2) {retrieve(dm2)};
\node[databox, below=of search2]                    (aresult2) {Diabetes Type II | ConceptID 477};
\node[innerbox, below=of result2]                   (abackend) {(Diabetes Type II AND Semaglutide) | CohortID: 77};

%%%%%%%%%%%%%%%%%%%%%%%%%%%%%%%%%%%%%%%%%%%%%%%%%%%%%%%%%%%%%%%%%%%%%%%%%%

% Define both database nodes
\node[data, aspect=0.25, draw, shape border rotate=90, minimum height=1, minimum width=2cm, below=1cm of processing,  fill=blue!10] (umls) {UMLS};
\node[data, aspect=0.25, draw, shape border rotate=90, minimum height=1, minimum width=2cm, below=0.25cm of umls, fill=red!17] (concept) {Concepts};

% Patient Data
\node[data, aspect=0.25, draw, shape border rotate=90, minimum height=1, minimum width=2cm, below=1cm of concept] (ehr) {EHR};

% Then the data layer grouping box that fits both
% \node[knowledgebox, fit=(umls) (concept), label=above:Knowledge] (data) {};
\node[knowledgebox, fit=(umls) (concept),
      label={[circle, draw, fill=white, font=\Large, scale=0.75, xshift=-2cm]above:4},
      label={[font=\small]above:Knowledge Base}] (data) {};

% TODO make invis box
\node[knowledgebox, draw=none, fit=(ehr), label=above:Patient Data] (patientdata) {};
\node[knowledgebox, draw=none, fit=(ehr)] (patientdata) {};

% Arrows
% \draw[arrow] (ehr.north) |- ++(-3.125, 1) |- (response.east);
% \draw[arrow] (retrieval2.east) -- (umls.west);
% \draw[arrow] (retrieval3.east) -- (umls.west);
% \draw[arrow] (retrieval4.east) |- ++(3, 0) -- (ehr.south);
% \draw[arrow] (backend.south) -- (ehr.north);

\draw[arrow] (interface.east) -- (processing);

% \draw[arrow] (processing.west) -- (retrieval.north east);
\draw[arrow] (search1.west) -- (retrieval.north east);
\draw[arrow] ([yshift=-19mm]search1.west) -- ([yshift=-19mm, xshift=1mm]retrieval.north east);

% \draw[arrow] (search2.west) -- (retrieval_stagger.north east);

% sub-agent dashed-lines
\draw[arrow] (retrieval2.east) -- (umls.west);
\draw[arrow] (retrieval6.east) -- (concept.west);

\draw[arrow] (response.east) --  ++(3mm,0) -- ++(0,-10cm) -- (ehr.west);

\end{tikzpicture}

%% file: figure_two.tex
\definecolor{nhsteal}{HTML}{005EB8}
\definecolor{sources_green}{RGB}{184,161,71}
\definecolor{ehs_gold}{RGB}{212,175,55}
\definecolor{internetpurple}{RGB}{90,60,160}

\begin{tikzpicture}[
    font=\huge,
    every label/.style={font=\huge\bfseries},
    node distance=1cm,
    outerbox/.style={rectangle, draw, thick, rounded corners, minimum width=1.5cm, minimum height=1.5cm, inner sep=5pt, line width=3pt},
    ehs/.style={
        outerbox,
        draw=ehs_gold,
        thick,
        rounded corners,
        minimum height=9cm,
        minimum width=6cm,
        align=center,
        inner sep=4mm,
        line width=3pt
    },
    group_users_and_sources/.style={
        outerbox,
        draw=sources_green,
        thick,
        rounded corners,
        minimum height=2mm,
        minimum width=1cm,
        align=center,
        inner sep=0mm,
        line width=3pt
    },
    internet/.style={
        ehs,
        draw=internetpurple,
        dash pattern=on 7pt off 7pt
    },
    database/.style={
        cylinder,
        shape border rotate=90,
        draw,
        fill=red!12,
        minimum height=2.5cm,
        minimum width=4cm,
        text width=3.5cm,
        align=center,
        aspect=0.4,
    },
    agent-box/.style={
        draw=yellow!17!black,
        fill=yellow!17,
        thick,
        rounded corners,
        minimum height=2cm,
        minimum width=4.8cm,
        align=center,
        inner sep=7mm
    },
    box-llm/.style={
        agent-box,
        draw=internetpurple,
        fill=blue!10,
        minimum height=1.7cm,
        minimum width=2cm,
        font=\huge\bfseries
    },
    icon-node/.style={
        thick,
        circle,
        minimum size=1.5cm,
        align=center,
        font=\Huge,
        scale=1.3
    },
    arrow_style/.style={
        line width=2.5pt
    },
    firewall/.style={
        draw=red,
        line width=4pt,
        dash pattern=on 7pt off 7pt
    },
]

% --- NODES ---
\node[icon-node, fill=white] (ui) {{\Large Frontend} \\\faDesktop};
\node[agent-box, right=of ui, xshift=2cm] (agent) {\Huge \textsc{ClinQueryAgent} \\ \\ \faCogs};

% EHR below UI, Concepts below Agent
\node[database, below=of ui, xshift=0cm, yshift=-6mm](patient_data) {EHR};
\node[database, below=of agent, yshift=-6mm] (concept_db) {Concepts};

% USERS
\node[icon-node, left=of ui, yshift=+5mm, xshift=-9cm] (phc_icon) {{\Large Practitioner} \\\faUserMd};
\node[icon-node, right=of phc_icon] (shc_icon) {{\Large Analyst} \\\faChartLine};
% \node[icon-node, right=of shc_icon] (ssc_icon) {{\Large Pharmacist} \\\faPills};
\node[icon-node, right=of shc_icon] (ssc_icon) {{\Large Commissioner} \\\faUserTie};

\node[group_users_and_sources, draw=nhsteal, fit=(shc_icon) (phc_icon) (ssc_icon), label={[color=nhsteal, label distance=3mm]left:Users}] (users) {};

% DATA SOURCES
\node[icon-node, below=of users, align=center, xshift=-5cm, yshift=-4mm] (hospitals)
    {\\[-0.3em]{\Large Hospitals}\\[0.2em]\faHospital};
\node[icon-node, right=of hospitals] (gps)
    {{\Large General} \\[-0.5em] {\Large Practices}\\\faClinicMedical};
\node[icon-node, right=of gps, align=center] (specialists)
    {{\Large Community} \\[-0.5em] {\Large Services}\\\faHome};

\node[group_users_and_sources, draw={rgb,255:red,94;green,184;blue,71}, fit=(hospitals) (gps) (specialists), align=center, label={[color={rgb,255:red,94;green,184;blue,71}, label distance=3mm]left:{
    \begin{tabular}{c}
    Data \\
    Sources
    \end{tabular}
}}] (sources) {};

% eHealthScope container — must come before firewall draw but after inner nodes
\node[ehs, fit=(ui) (agent) (patient_data) (concept_db), label={[color={rgb,255:red,184;green,161;blue,71}] NHS eHealthScope}] (ehealth) {};

% Internet
\node[box-llm, right=of agent, xshift=2cm, yshift=-3.5mm] (llm) {LLM};
\node[box-llm, below=of llm] (umls) {UMLS};

\node[internet, fit=(llm) (umls), label={[internetpurple]:Internet}] (www_fit) {};

\coordinate (fw_mid) at ($(agent.south west)!0.35!(patient_data.north east)$);

\draw[firewall] 
    ($(fw_mid)+(0,4.3cm)$) 
    -- 
    ($(fw_mid)+(0,-4.1cm)$);
\node[fill=white, inner sep=2pt, text=red, font=\Huge, anchor=north] 
    at ($(fw_mid)+(0,-4.5cm)$)
    {\faLock~\textbf{Firewall}};

% --- ARROWS ---
\draw[->, arrow_style] 
    (users.east |- {$(users.east)!0.5!(ui.west)$}) 
    -- 
    (ui.west |- {$(users.east)!0.5!(ui.west)$});\draw[->, arrow_style] (agent) to (concept_db);
\draw[->, arrow_style] (agent) to (umls);
\draw[->, arrow_style] (agent.east) to (llm.west);
\draw[->, arrow_style] (patient_data) to ([yshift=+1.5em]ui.south);
\draw[->, arrow_style] (ui) to (agent);

% \draw[<->, arrow_style] (sources) to (patient_data);
\draw[->, arrow_style] 
    (sources.east |- {$(sources.east)!0.5!(patient_data.west)$}) 
    -- 
    (patient_data.west |- {$(sources.east)!0.5!(patient_data.west)$});
\end{tikzpicture}

%% file: figure_three.tex
\begin{tikzpicture}[
    node distance=1mm, % vertical spacing between nodes
    % --- Styles imported from Reference ---
    font=\small,
    innerbox/.style={rectangle, draw, solid, thick, rounded corners, minimum width=1.5cm, minimum height=0.5cm, align=left, text width=3cm, font=\small},
    outerbox/.style={rectangle, draw, thick, rounded corners, minimum width=3.5cm, minimum height=2cm, inner sep=4pt},
    knowledgebox/.style={outerbox, minimum width=2.5cm, minimum height=0.5cm},
    promptbox/.style={innerbox, fill=yellow!17},
    databox/.style={rectangle, draw, solid, minimum width=2.8cm, text width=2.8cm, minimum height=0.5cm, align=left, font=\small, fill=white, align=left},
    retrievetool/.style={rectangle, draw, solid, minimum width=2.8cm, text width=2.8cm, minimum height=0.5cm, align=left, font=\small, fill=teal!7},
    searchtool/.style={rectangle, draw, solid, minimum width=2.8cm, text width=2.8cm, minimum height=0.5cm, align=left, font=\small, fill=red!17},
    umlstool/.style={rectangle, draw, solid, minimum width=2.8cm, text width=2.8cm, minimum height=0.5cm, align=left, font=\small, fill=blue!10},
    linkbox/.style={innerbox, fill=yellow!17},
    arrow/.style={-Stealth, thick},
    arrowdashed/.style={-Stealth, thick, dashed},
    data/.style={cylinder, font=\small}
]

% ==========================================================================================
% 1. QUERY AGENT
% ==========================================================================================

% --- User Input ---
% Style: innerbox, fill=magenta!17 (matches User Interface component from reference)
\node[innerbox, fill=orange!17] (user_msg) at (0,0) 
    {"What's the rate of people with \textbf{dm2} who take \textbf{ozempic} in each district?"};

% --- Action 1 ---
% Style: retrievetool (matches 'retrieve' nodes from reference)
\node[retrievetool, below=of user_msg] (q_act1) 
    {retrieve(dm2)};

% --- Context 1 ---
% Style: databox (matches result nodes from reference)
\node[databox, below=of q_act1] (q_ctx1) 
    {$\bullet$ Diabetes II};

% --- Action 2 ---
\node[retrievetool, below=of q_ctx1] (q_act2) 
    {retrieve(semaglutide)};

% --- Context 2 ---
\node[databox, below=of q_act2] (q_ctx2) 
    {$\bullet$ Semaglutide last 12mos\\
    $\bullet$ Semaglutide last 3mos\\};

% --- Action 3 ---
% Style: innerbox (Generic processing step)
\node[innerbox, below=of q_ctx2] (q_act3) 
    {create\_cohort(\\
    \phantom{x} "Diabetes II AND\\
    \phantom{x} Semaglutide last 3mos"\\)};

% --- Final Return ---
% Style: linkbox (Matches response/link nodes from reference)
\node[linkbox, below=of q_act3] (q_final) 
    {return\_url(\\ \hspace{1em}/register?ao=77\\)};

% ==========================================================================================
% 2. RETRIEVAL AGENT 1 (DM2)
% ==========================================================================================

% Style: searchtool (matches 'search' nodes)
\node[searchtool, right=1cm of q_act1, yshift=22mm] (s1_act1) 
    {search(dm2)};

% Style: databox (results)
\node[databox, below=of s1_act1] (s1_fail) 
    {[ 0 results found ]};

\node[searchtool, below=of s1_fail] (s1_act2) 
    {search(diabetes II)};

\node[databox, below=of s1_act2] (s1_res) 
    {$\bullet$ Diabetes II\\
     $\bullet$ Diabetes II resolved\\
      \textit{\textbf{+405 additional concepts}}};

% Style: databox (return value)
\node[databox, below=of s1_res, fill=teal!7] (s1_ret) 
    {
    return\_shortlist([\\
    \hspace{2em}$\bullet$ Diabetes II\\
    ])};

% ==========================================================================================
% 3. RETRIEVAL AGENT 2 (SEMAGLUTIDE)
% ==========================================================================================

\node[searchtool, below=1.0cm of s1_ret, xshift=0cm] (s2_act1) 
    {search(ozempic)};

\node[databox, below=of s2_act1] (s2_fail) 
    {[ 0 results found ]};

% Style: umlstool (matches 'lookup' nodes)
\node[umlstool, below=of s2_fail] (s2_lookup) 
    {lookup(ozempic)};

\node[databox, below=of s2_lookup] (s2_found) 
    {$\bullet$ Ozempic \\ \hspace{1em}(Semaglutide)};

\node[searchtool, below=of s2_found] (s2_act2) 
    {search(semaglutide)};

\node[databox, below=of s2_act2] (s2_res) 
    {$\bullet$ Semaglutide last 3mos\\
     $\bullet$ Semaglutide last 12mos\\
     $\bullet$ Semaglutide \\Eligible\\
     % \textit{+ additional results}
     };

\node[databox, below=of s2_res, fill=teal!7] (s2_ret) 
    {
    return\_shortlist([\\
    \hspace{2em}$\bullet$ Semaglutide \\\hspace{3em}last 3mos\\
    \hspace{2em}$\bullet$ Semaglutide \\\hspace{3em}last 12mos\\
    ])};

% ==========================================================================================
% 4. USER VIEW (IMAGE PANE)
% ==========================================================================================

% Keeping structural placement, using innerbox style (white fill implied or set)
\node[
    rectangle, draw, thick, rounded corners, minimum width=3.5cm, inner sep=0pt,
    below=6.6mm of q_final,
    label={[font=\small]south:eHealthScope GUI}
] (user_view_pane) {
    \includegraphics[width=4cm]{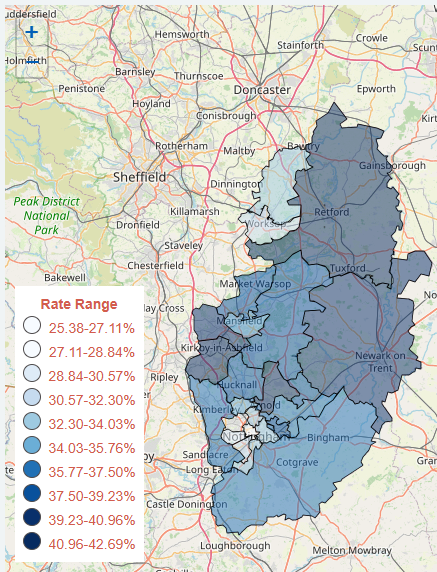}
};

% Arrow from URL return to User Interface
\draw[arrowdashed] (q_final.south) -- (user_view_pane.north);

% ==========================================================================================
% BACKGROUND BOXES (drawn on background layer)
% ==========================================================================================

\begin{pgfonlayer}{background}
    % Query Agent Background
    % Style: outerbox, fill=yellow!17, labeled like reference (Circle 1, Text)
    \node[outerbox, fill=yellow!17, fit={(user_msg) (q_final)}, 
          label={[font=\small]above:Query Agent}] (main_agent_box) {};
    
    % RA 1 Background
    % Style: outerbox, fill=teal!7
    \node[outerbox, fill=teal!7, fit={(s1_act1) (s1_ret)}, 
          label={[font=\small]above:{\hspace{1em}Retrieval Agent 1}}] (sub1_box) {};
    
    % RA 2 Background
    % Style: outerbox, fill=teal!7
    \node[outerbox, fill=teal!7, fit={(s2_act1) (s2_ret)}, 
          label={[font=\small]above:{\hspace{1em}Retrieval Agent 2}}] (sub2_box) {};
\end{pgfonlayer}

% ==========================================================================================
% 5. CONNECTIONS
% ==========================================================================================

% Flow 1: Call (QA -> RA1) - angled
% \draw[arrow] (q_act1.east) -- ++(0.4,0) |- (s1_act1.west);
\draw[arrow] (q_act1.east) -- ++(0.5,0) |- (sub1_box.north west);

% Flow 1: Return (RA1 -> QA) - angled
% \draw[arrowdashed] (s1_ret.west) -- ++(-0.4,0) |- (q_ctx1.east);

% Flow 2: Call (QA -> RA2) - angled
% \draw[arrow] (q_act2.east) -- ++(0.4,0) |- (s2_act1.west);
\draw[arrow] (q_act2.east) -- ++(0.5,0) |- (sub2_box.north west);

% Flow 2: Return (RA2 -> QA) - angled
% \draw[arrowdashed] (s2_ret.west) -- ++(-0.4,0) |- (q_ctx2.east);

\end{tikzpicture}